\documentclass[12pt,a4paper]{article}
\usepackage[utf8]{inputenc}
\usepackage[margin=1in]{geometry}
\usepackage{amsmath} 
\usepackage{amsfonts}
\usepackage{amssymb} 
\usepackage{lineno}
\usepackage{fullpage}
\usepackage{setspace}
\usepackage{graphicx}
\usepackage{float} 
\usepackage{breakcites}
\usepackage{pdfpages}
\usepackage{xcolor}
\usepackage{multicol}
\usepackage{natbib}
\usepackage{hyperref}
\usepackage[ruled,vlined]{algorithm2e}

\DeclareMathOperator*{\argmin}{argmin}

\makeatletter
\def\BState{\State\hskip-\ALG@thistlm}
\makeatother

\title{A Fast Particle-Based Approach for Calibrating a 3-D Model of the Antarctic Ice Sheet}
\author{Ben Seiyon Lee, Murali Haran, Robert Fuller, David Pollard, and Klaus Keller}

\graphicspath{{Figures/}}

\begin{document}

\maketitle

        \begin{abstract}
We consider the scientifically challenging and policy-relevant task of understanding the past and projecting the future dynamics of the Antarctic ice sheet.  The Antarctic ice sheet has shown a highly nonlinear threshold response to past climate forcings. Triggering such a threshold response through anthropogenic greenhouse gas emissions would drive drastic and potentially fast sea level rise with important implications for coastal flood risks. Previous studies have combined information from ice sheet models and observations to calibrate model parameters. These studies have broken important new ground but have either adopted simple ice sheet models or have limited the number of parameters to allow for the use of more complex models. These limitations are largely due to the computational challenges posed by calibration as models become more computationally intensive or when the number of parameters increases. 

Here we propose a method to alleviate this problem: a fast sequential Monte Carlo method that takes advantage of the massive parallelization afforded by modern high performance computing systems. We use simulated examples to demonstrate how our sample-based approach provides accurate approximations to the posterior distributions of the calibrated parameters. The drastic reduction in computational times enables us to provide new insights into important scientific questions, for example, the impact of Pliocene era data and prior parameter information on sea level projections. These studies would be computationally prohibitive with other computational approaches for calibration such as Markov chain Monte Carlo or emulation-based methods. We also find considerable differences in the distributions of sea level projections when we account for a larger number of uncertain parameters. For example, based on the same ice sheet model and data set, the 99th percentile of the Antarctic ice sheet contribution to sea level rise in 2300 increases from 6.5 m to 13.1 m when we increase the number of calibrated parameters from three to 11. With previous calibration methods, it would be challenging to go beyond five parameters. This work provides an important next step towards improving the uncertainty quantification of complex, computationally intensive, and decision-relevant models. 

    \end{abstract}

\section{Introduction}\label{Sec:Intro}
How much will the Antarctic ice sheet contribute to future sea level rise? 
The geological records suggest that ice sheets can quickly contribute considerable amounts to global sea level rise \citep{deschamps2012ice}, in some cases up to 58 m \citep{fretwell2012bedmap2}.
Projections of future sea level rise depend on deeply uncertain projections of the Antarctic ice sheet's (AIS) mass loss \citep{le2017high,wong2017impacts,le2017bounding}. Close to eight percent of the current global population is threatened by a five meter rise in sea level \citep{nicholls2008global} and 13 percent of the global urban population is threatened by a ten meter sea level rise \citep{mcgranahan2007rising}. Quantifying and characterizing the long-term behavior of the Antarctic ice sheet is hence a key input to the design of coastal risk management strategies \citep[cf.][]{garner2018using,sriver2018characterizing,oppenheimer2016high}.

Ice sheet models rely on poorly constrained parameters, and recent studies show that uncertainty in model parameters results in highly uncertain projections of sea level change \citep{stone2010investigating,applegate2012assessment,fitzgerald2012exploration,collins2007ensembles}; thereby  affecting climate risk decision-making \citep{oneill2006learning,hannart2013disconcerting}. Recent studies have addressed parametric uncertainty via calibration studies using modern observations, but these are either limited to simple ice sheet models \citep{ruckert2017assessing,fuller2017probabilistic} or a small number of model parameters \citep{chang2016improving,Edwards2019,Schlegel2018}. Numeric solvers have been used to infer the field of basal sliding parameters from satellite observations \citep{isaac2015solution, isaac2015scalable}.

Ice sheet models vary in complexity, and the key drivers of computational cost are the spatial and temporal resolutions. Simpler models \citep[cf.][]{shaffer2014formulation,bakker2016simple} have short computer model run times on the order of a few seconds, but they may oversimplify or even exclude important physical processes. More complex models \citep[cf.][]{deconto2016contribution,larour2012continental,greve1997application,rutt2009glimmer} can better represent key ice dynamics and typically run at higher spatio-temporal resolutions. However, they require longer model run times. Here, we use a relatively complex ice sheet model, the Pennsylvania State University 3D ice sheet model (PSU3D-ICE) \citep{pollard2012description}, but with considerably coarser resolution than in previous work, so that each set of simulations for this study takes on the order of 10 to 15 minutes of wall time. 

Past studies calibrate simpler models with many model parameters using Markov Chain Monte Carlo (MCMC) \citep[cf.][]{ruckert2017assessing,bakker2016simple, petra2014computational}; these approaches are effective in the context of computationally very inexpensive models (model run times of a few seconds), and hence do not extend to the kind of models we consider in this manuscript. Some studies have employed emulation-calibration methods \citep{sanso2008inferring,Liu2009Mod,bhat2010computer} 
to calibrate computer models with long run times, but these approaches are applicable to only a small number of parameters. For computer models with longer run times and a large number of model parameters, emulation-calibration can be computationally prohibitive because building an accurate emulator requires a very large set of training data \citep{bastos2009diagnostics,maniyar2007dimensionality}. 

We propose calibrating an ice sheet model which (1) accounts for important physical processes; (2) includes several key parameters to analyze and quantify parametric uncertainty; and (3) expands the calibration dataset to the Pliocene. For this study, the Antarctic ice sheet model runs at a spatial resolution of $80$ km and temporal resolution of eight years, which is a compromise between preserving reasonable accuracy of physical simulations versus maintaining a feasible model run time. We estimate that current rigorous methods for calibrating this model via Markov chain Monte Carlo would take roughly on the order of years of wall time. We investigate methods for calibration that are amenable to heavy parallelization and computationally efficient, thereby reducing the computational wall time from years to hours. We find that these methods are broadly applicable to computer models with a moderate model run time (6 seconds to 15 minutes) and a moderate number of model parameters (5 to 20), based on available computing resources. While this does not cover more complex models or larger number of parameters, our methods are applicable to many scientifically important and policy-relevant computer models.

Studying the Antarctic ice sheet's future behavior motivates the need for a computationally efficient approach for computer model calibration. We turn to sequential Monte Carlo methods \citep[cf.][]{doucet2000sequential,del2006sequential,chopin2002sequential}, building upon particle-based methods for computer model calibration \citep{higdon2008computer,kalyanaraman2016uncertainty}. Our approach builds upon an adaptive tempering schedule and an adaptive mutation stage \citep{jasra2011inference}, which have been used for Bayesian variable selection \citep{schafer2013sequential}, Bayesian model comparison \citep{zhou2016toward}, and estimating initial conditions of the Navier-Stokes system of equations \citep{kantas2014sequential,llopis2018particle}.

By using massive parallelization in a high performance computing environment, we obtain a dramatic speed-up over current MCMC-based calibration methods, roughly reducing wall time by a factor of 3000. 
We also limit expensive computer model runs by imposing stopping rules and adaptive sampling techniques. 
We provide practical guidelines designed to: (1) reduce total wall time; (2) limit the number of expensive computer model runs; and (3) simplify implementation for the user. Our computationally efficient calibration approach is readily applicable to many computer models for which rigorous calibration may be currently infeasible. 

We note that we focus on a `static' system where all observations are available at once; hence, there is only one posterior distribution of interest, which we approximate using our particle-based approach. The PSU3D-ICE model is dissipative where it evolves to a single constant steady state for a given set of parameter values and external forcing \citep{willems1972dissipative}. Unlike choatic systems such as global weather models, ``microscopic" changes in the initial states do not change the results; in other words, there is no ``butterfly effect'' \citep{lorenz1972}. We use our approach to calibrate the PSU3D-ICE model \citep{deconto2016contribution} using paleoclimate data and modern observational records. Previous work focuses on calibrating the PSU3D-ICE model using fewer parameters \citep{chang2016improving,Edwards2019} or surrogate models using limited training data \citep{chang2016calibrating}. 
Using our new method, we show that the information regarding the extent of the Antarctic ice sheet in the Pliocene era strongly influences parametric and projection uncertainty. We find that using improved geological data and analysis to characterize the Antarctic ice sheet's contribution to sea level rise in the Pliocene can bring about considerably sharper sea level projections for future centuries. 

    The paper is structured as follows. In Section \ref{Sec:PSU3DICE}, we provide an  overview of the ice sheet model (PSU3D-ICE). In Section \ref{Sec:Calibration}, we describe the model calibration framework and discuss challenges with current calibration methods. We propose our fast particle-based approach for computer model calibration in Section \ref{Sec:OurApproach}. In Section \ref{Sec:Simulation}, we demonstrate the application of our method to a simulated example. In Section \ref{Sec:Results} we apply our method to the PSU3D-ICE model and report our scientific conclusions. We end with caveats and directions for future research in Section \ref{Sec:Discussion}. 

\section{Computer model description and observations}\label{Sec:PSU3DICE}
In this section, we provide background information for the PSU3D-ICE Antarctic ice sheet model \citep{deconto2016contribution} as well as the paleoclimate records and modern observations used to calibrate the model. 

\subsection{The PSU3D-ICE model}
The PSU3D-ICE model simulates the long-term dynamics of continental ice sheets. It has previously been applied to past and future variations of the Antarctic ice sheet \citep{pollard2009modelling, pollard2012description, pollard2015potential, Pollard2016Large, pollard2017variations}. Slow ice deformation under its own weight is modeled by scaled dynamical equations for internal shear, horizontal stretching, and basal sliding. Other variables and processes include internal ice temperatures, bedrock deformation beneath the ice load, surface snowfall and melting, oceanic melting beneath floating ice shelves, and calving of ice into the ocean \citep{pollard2012description}. A recently proposed mechanism called Marine Ice Cliff Instability (MICI) that can drastically attack ice in marine basins, involving hydrofracturing due to surface liquid water and structural failure of tall ice cliffs, is included here \citep{pollard2015potential, deconto2016contribution}. Note that this mechanism has recently been questioned \citep{Edwards2019, Golledge2019}.   

For the simulations in this study, a polar stereographic grid spanning Antarctica is used with a horizontal resolution of 80 kilometers (km), which yields a model run time of approximately 10 to 15 minutes for each set of past and future simulations described below. This is a considerably coarser spatial resolution than previous continental-scale applications, which have used resolutions of 10 to 40 km. However, sensitivity tests with the model show reasonable independence of results with model resolution, due to the grid-independent parameterization of important sub-grid processes such as grounding-line flux and cliff failure \citep{pollard2015potential}. Those tests and the reasonable agreement in additional limited offline tests at 80 km vs. finer resolutions indicate that the coarser resolution is adequate for this study.

\begin{figure}[h]
    \centering
    \includegraphics[width=1\textwidth]{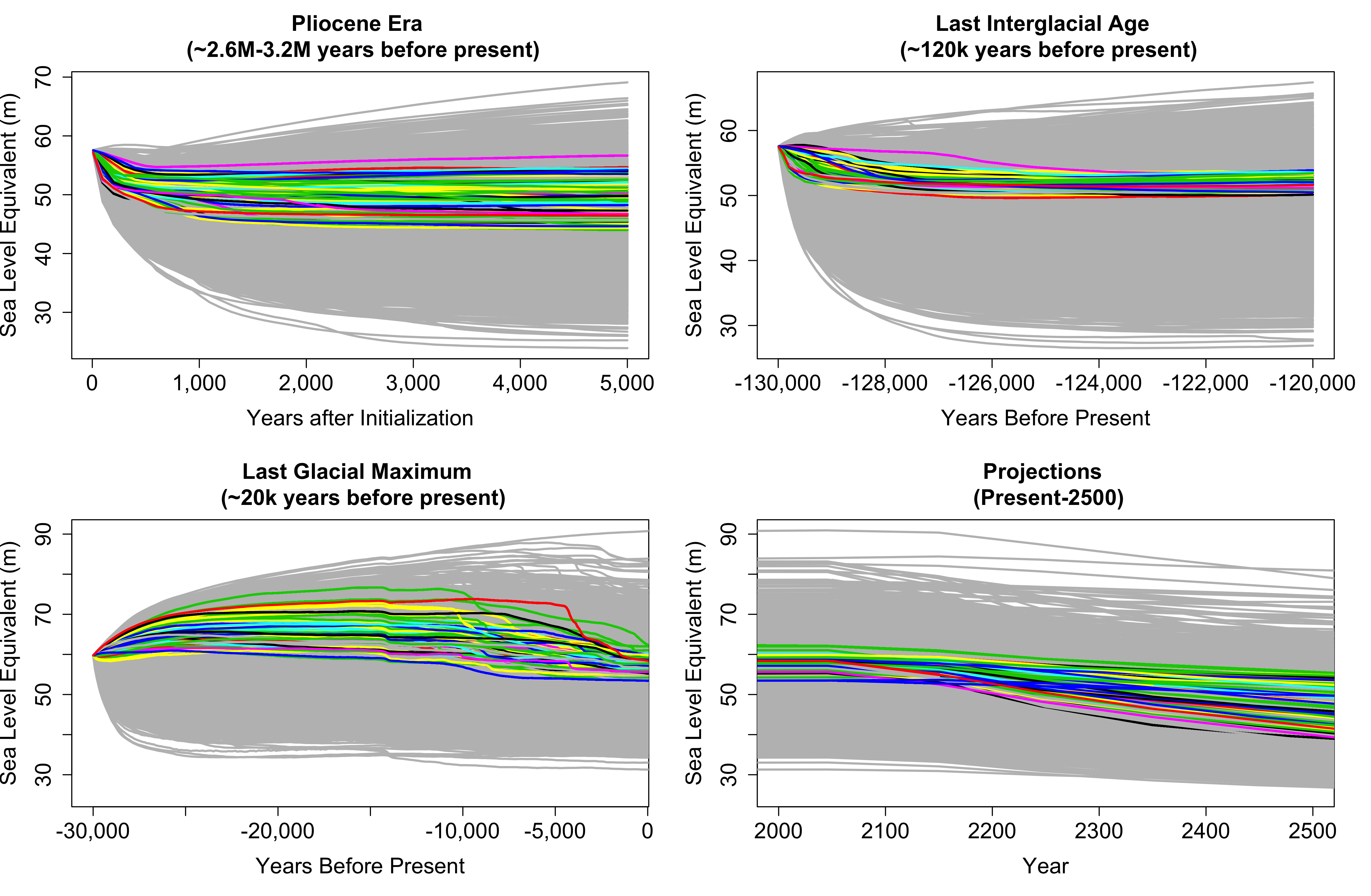}
    \caption{Time series of 1500 simulated model output from the PSU3D-ICE model where each model run corresponds to a line. Data are generated using 1500 parameter sets from the prior distribution. The y-axis denotes the Antarctic ice sheet's contribution to sea level change in meters (m). We approximate the present as year 1950. Model simulations that have a non-zero likelihood are denoted by colored lines and runs that have a zero likelihood are displayed in gray. (Top left) Model output for the Pliocene era model run where the x-axis denotes years after initialization. (Top right) Model output for the Last Interglacial Age where the x-axis denotes years before the present. (Bottom left) Model output for the Last Glacial Maximum where the x-axis denotes years before the present. (Bottom right) Model projections for 2000-2500 where the x-axis represents years.}
    \label{SuppFig:ModelRuns}
\end{figure}

We evaluate the PSU3D-ICE model over three separate time periods. As in previous ensemble work with this model \citep{chang2016calibrating, chang2016improving, Pollard2016Large, deconto2016contribution}, the time periods are selected to include major ice-sheet variations that stringently test the model and have at least some paleo data to provide useful quantitative constraints. The three time periods are: (1) a single episode of high sea level rise during the warm mid-Pliocene (which extended roughly from ~3.2 to 2.6 million years before present); (2) the Last Interglacial period around 125,000 to 115,000 years ago, at the start of the last Pleistocene glacial-interglacial cycle when global climate was slightly warmer than today, the major Northern Hemispheric ice sheets were most recently absent prior to the modern interglacial period, Greenland was smaller, and the West Antarctic ice sheet may have undergone major collapse; and (3) the last deglacial period from the Last Glacial Maximum  about 20,000 years ago to the present, and then 5,000 years into a warmer future. In Figure \ref{SuppFig:ModelRuns}, we present 1500 model simulations from the PSU3D-ICE model for all three time periods and projections into the future. We describe the three model simulations below.

To represent a single high sea level episode during the  warm mid-Pliocene era (roughly 3.2 to 2.6 million years before present), we initialize the ice sheet model to modern conditions and run the model forward for 5,000 years. As described in previous Pliocene applications \citep{pollard2015potential, pollard2017variations}, atmospheric climatic forcing is provided by the RegCM3 regional climate model \citep{pal2007regional} adapted for polar regions, driven by the GENESIS v3 global climate model \citep{Alder2011}. The atmospheric carbon dioxide concentration is set to at 400 parts per million by volume (ppmv), and a warm austral summer orbit is specified. We use oceanic temperatures from the modern World Ocean Atlas database \citep{levitus2012world}, with a +2 $^{\circ}$C uniform perturbation added to represent mid-Pliocene ocean warming. Atmospheric monthly cycles of surface air temperature and precipitation are used to compute melting and annual mass balance on the ice-sheet surface, and oceanic temperatures are used to compute basal melting under floating ice shelves \citep{pollard2015potential}.

For the Last Interglacial (LIG), we initialize the ice sheet model to modern conditions and run the model from 130,000 to 120,000 years before present (130 ka to 120 ka). As described in \citet{deconto2016contribution}, LIG climates are specified as uniform perturbations to modern climatology (\citet{le2010improved} for atmosphere,  and \citet{levitus2012world} for ocean). The atmospheric and ocean temperature perturbations vary step-wise in time. From 130 ka to 125 ka, they are $+1.97^{\circ}$C and $+1.70^{\circ}$C respectively. From 125 ka to 120 ka, they are $+1.41^{\circ}$C and $+1.51^{\circ}$C respectively.

The Last Glacial Maximum, modern, and future eras are simulated in one continuous run, over the last 40,000 years through the last deglacial period to modern, and extended 5,000 years into the future. As described in \citet{Pollard2016Large},  the model is initialized appropriately at 40 ka (40,000 years Before Present or BP, relative to 1950 AD) from a previous long-term run. Atmospheric forcing is supplied using a modern climatological Antarctic dataset \citep{le2010improved}, with uniform cooling perturbations applied proportional to a deep sea-core $\delta^{18}$O record \citep{pollard2009modelling,pollard2012description}. Oceanic forcing is supplied from a coupled Atmosphere-Ocean General Circulation Model (AOGCM) simulation of the last 20,000 years \citep{liu2009transient}. After reaching present day, each run is extended for 5,000 years with atmospheric and oceanic forcing as described in \citet{deconto2016contribution}, for the Representative Concentration Pathway (RCP) 8.5 scenario of future greenhouse gas emissions and concentrations \citep{meinshausen2011rcp}, often called `business as usual'. Atmospheric temperatures and precipitation are obtained by appropriately weighting previously saved simulations of the RegCM3 regional climate model for particular carbon dioxide levels, and oceanic temperatures are supplied from an archived transient NCAR global model simulation \citep{shields2016atmospheric}. 

After each model run, we extract the pertinent model output, specifically the Antarctic ice sheet's contribution to sea level change ($m$), total ice volume ($km^{3}$), and total grounded ice area ($km^{2}$). We then compare this to the corresponding paleo- or modern observational records. In this study, we examine 11 model parameters considered to be important in modeling the behavior of the Antarctic ice sheet - OCFACMULT, OCFACMULTASE, CRHSHELF, CRHFAC, ENHANCESHEET, ENHANCESHELF, FACEME-LTRATE, TAUASTH, CLIFFVMAX, CALVLIQ, and CALVNICK. Detailed descriptions of each parameter are provided in the Appendix. 

We note that this is a much larger number of parameters than typically considered for models with such detailed dynamics. The ice sheet model has many more parameters than the 11 chosen here. The values for many of them are reasonably well established in the glaciological literature, resulting from published work over the last several decades applying similar models to the Antarctic ice sheet. Those parameters mostly involve terrestrial processes (i.e., where ice is grounded on bedrock) that are constrained directly or indirectly by observational data of the modern ice sheet, and/or laboratory ice physics, such as the rheology of ice, ice streaming vs. shearing flow, basal sliding coefficients, and modern ice distribution and thicknesses. The 11 parameters chosen here can have large effects on the results, but are not well constrained by modern observations because they apply to processes (1) that have occurred in the past and expected in the future, but are not active today, or (2) are undergoing rapid change in recent decades.  Examples of (1) are basal sliding coefficients for bedrock in modern ocean regions where grounded ice advanced during past glacial maxima, and the timescale of bedrock rebound under varying ice loads. Examples of (2) are coefficients for oceanic melting at the base of floating ice shelves, and oceanic melting at vertical ice fronts. A subset of these parameters have been used in more limited ensembles with this model \citep{chang2016improving, chang2016calibrating, Pollard2016Large, pollard2017variations}, but here the 11 parameters constitute the bulk of important yet relatively unconstrained parameters in the model.

\subsection{Paleoclimate records and modern observations}\label{Sec:Observations}
For the paleoclimate records, we use the Antarctic ice sheet's contribution to sea level change in the following eras: Pliocene ($\sim$2.6-3.2 million years before present); the Last Interglacial Age ($\sim$125,000 to 115,000 years before present); and the Last Glacial Maximum ($\sim$20,000 years before present). We specify the Antarctic ice sheet's contribution to sea level change in terms of global mean sea level equivalents (SLE) relative to the modern ice sheet, thereby correctly allowing for marine ice grounded below sea level. The base units are meters ($m$). We adopt the following ranges for the paleoclimate records, which account for considerable uncertainty in published estimates \citep[cf.][]{kopp2009probabilistic, dutton2015sea}: (1) 5 m to 25 m for the Pliocene \citep{naish2009obliquity,rovere2014mid,cook_dynamic_2013}; (2) 3.5 m to 7.5 m for the Last Interglacial Age \citep{fuller2017probabilistic,deconto2016contribution}; and (3) -5 m to -15 m for the Last Glacial Maximum \citep{ruckert2017assessing,Pollard2016Large}. 

Modern observations include total volume and grounded area of the Antarctic ice sheet, as well as ten spatial locations that currently have ice present. Units for total volume and total grounded ice area are cubic kilometers ($km^3$) and square kilometers ($km^2$) respectively. Observations come from the Bedmap2 dataset \citep{fretwell2012bedmap2}, which provide the most recent gridded maps of ice surface elevation, bedrock elevation, and ice thickness. The Bedmap2 maps are generated using multiple sources, including satellite altimetry, airborne and ground radar surveys, and seismic sounding.

\section{Model calibration framework}\label{Sec:Calibration}
In this section, we describe the general computer model calibration framework. In computer model calibration, key computer model parameters are estimated by comparing the computer model output and observational data \citep[cf.][]{chang2016calibrating,kennedy2001bayesian,bayarri2007computer,bhat2010computer}. Calibration methods also account for key sources of uncertainty such as model-observation discrepancy and observational error \citep{kennedy2001bayesian,bayarri2007computer,brynjarsdottir2014learning}. We describe a model for output in the form of spatial data as this directly relates to our simulated data example in Section \ref{Sec:Simulation}; a time series version of this applies to the PSU3D-ICE model in Section \ref{Sec:Results}.

Let $Y(s,\theta)$ be the computer model output at the spatial location $s \in \mathcal{S} \subseteq \mathbb{R}^{2}$ and the parameter setting $\theta\in \Theta \subseteq \mathbb{R}^{d}$. $\mathcal{S}$ is the spatial domain of the process, and $\Theta$ is the parameter space of the computer model with integer $d$ being the number of input parameters. $\mathbf{Y}=(Y(s_{1},\theta_{i}),...,Y(s_{n},\theta_{i}))^{T}$ is the computer model output at parameter setting $\theta_{i}$ and spatial locations $(s_{1},...,s_{n})$. $\mathbf{Z}=(Z(s_{1}),...,Z(s_{n}))^{T}$ is the observed spatial process at locations $(s_{1},...,s_{n})$. 

We model the observational data $Z$ as follows,
\begin{equation}\label{EQ:ModelCalibration}
  Z=Y(\theta) + \delta + \epsilon,  
\end{equation}
where $\epsilon\sim N(0,\sigma_{\epsilon}^{2})$ is independently and identically distributed observational error, and $\delta$ is a systemic data-model discrepancy term. The discrepancy $\delta$ is modeled as a zero-mean Gaussian process, where $\delta\sim N(0,\Sigma_{\delta}(\xi_{\delta}))$. This discrepancy term is essential for parameter calibration \citep{bhat2010computer,bayarri2007computer} and ignoring it may yield biased and overconfident estimates and projections \citep{brynjarsdottir2014learning}. $\Sigma_{\delta}(\xi_{\delta})$ is the spatial covariance matrix between spatial points $s_{1},...,s_{n}$ with covariance parameters $\xi_{\delta}$. We set standard prior distributions for the model parameters, $\theta$, and observational error variance, $\sigma_{\epsilon}^{2}$. On the other hand, informative priors are necessary for the discrepancy term's covariance parameters $\xi_{\delta}$. Then, we infer $\theta$, $\sigma^{2}$, and $\xi_{\delta}$ by sampling from the posterior distribution, $\pi(\theta,\sigma_{\epsilon}^{2},\xi_{\delta}|Z)$, via Markov Chain Monte Carlo (MCMC).

\subsection*{Challenges with computer Model Calibration}
We focus on a specific class of computer models, characterized by (1) a moderate run time (6 seconds to 15 minutes); and (2) moderately large parameter space (5 to 20 parameters). The modified PSU3D-ICE Antarctic ice sheet model (Section \ref{Sec:PSU3DICE}) fits the specifications for this class of computer models. Several other important models that can potentially be modified to fit within this class are single column atmospheric models \citep{bony2001parameterization,Dal2018,Gettelman2019}, simplified earth systems models \citep{monier2013integrated}, hydrological soil moisture models \citep{Sorooshian1993, Liang1994}, and integrated multi-Sector models for human and earth dynamics \citep{kim2006bj}. 

The calibration framework requires running the computer model once for each iteration of the MCMC algorithm. Subject to overall calibration wall times, MCMC-based calibration methods are well suited to computer models that run very quickly, typically under 6 seconds per model run. The PSU3D-ICE model takes approximately 10 to 15 minutes per run on a single 2.3-GHz Intel Xeon E5-2697V4 (Broadwell) processor. We estimate that a standard MCMC-based calibration approach for this would take on the order of 2.9 years to approximate the posterior distribution $\pi(\theta|Z)$. 

Surrogate methods such as Gaussian process-based emulators are well suited to computer models with long run times. A good design is important for building accurate surrogates. Dense sampling schemes, such as full factorial or fractional factorial designs, capture higher order interactions; however, running the computer model at each of the design points is costly. Space-filling designs such as the Latin Hypercube Design \citep{mckay2000comparison,steinberg2006construction,stein1987large} or adaptive experimental designs \citep{chang2016calibrating,gramacy2015local,urban2010comparison,queipo2005surrogate} use fewer design points, but may possibly generate low-fidelity surrogate models by ignoring higher order interactions among inputs \citep{liu2017dimension}. Since the PSU3D-ICE model exhibits non-linear dependencies among input parameters \citep{pollard2012description}, we would be limited to 6 or fewer parameters using standard emulation-calibration techniques (with our available computing resources). 

\section{Fast particle-based calibration}\label{Sec:OurApproach}
In this section, we present a fast particle-based method to calibrate computers models with moderate model run time (6 seconds to 15 minutes) and a moderate number of model parameters (5 to 20). We begin with a description of a sequential sampling-importance-resampling algorithm. Then, we present modifications to the algorithm designed to improve computational efficiency. We examine advantages and limitations of our approach. Finally, we discuss tuning mechanisms for our method and provide practical guidelines.

\subsection{Sequential sampling-importance-resampling with mutation}
We propose a series of sampling-importance-resampling with mutation operations, which includes evolving importance and target distributions. The objective is to efficiently approximate a target distribution using a swarm of evolving particles. Our approach falls under the umbrella of Sequential Monte Carlo algorithms \citep{del2006sequential,doucet2000sequential,liu2001combined}, which have gained wide practical use \citep[cf.][]{kantas2015particle,papaioannou2016sequential,kalyanaraman2016uncertainty,jeremiah2011bayesian,morzfeld2018iterative}. In particular, we build upon the Iterated Batch Importance Sampling (IBIS) \citep{chopin2002sequential,crisan2000convergence} method. 

\subsubsection*{Sampling-importance-resampling}
Sampling-Importance-Resampling \citep{gordon1993novel,doucet2001introduction} is a sampling method used to approximate a target distribution $\pi(\theta)$ using samples from an importance distribution $q(\theta)$. Suppose we want to estimate $\mu=E_{\pi}\big[g(\theta)\big]$. Given $q(\theta)>0$ whenever $g(\theta)\pi(\theta)>0, \quad \forall \theta \in \Theta$, we observe that $E_{\pi}\big[g(\theta)\big]=E_{q}\Big[g(\theta)w(\theta)\Big]$, where $w(\theta)=\frac{\pi(\theta)}{q(\theta)}$ is the importance weight and $\sum_{i=1}^{N}w(\theta_{i})=1$. The importance sampling estimator is $\hat{\mu}_{n}=\frac{1}{n}\sum_{i=1}^{N}g(\theta_{i})w(\theta_{i})$ and $\hat{\mu}_{n}\rightarrow \mu$ with probability 1 by the strong law of large numbers. For target distributions with an unknown normalizing constant, such as the posterior distribution of the model calibration parameters $\pi(\theta|Z)$, the importance weights $w(\theta_{i})$, must be normalized.

An extension of importance sampling is sampling-importance-resampling, which provides an approximation of a target distribution via samples from an importance distribution and corresponding importance weights \citep{gordon1993novel}. The target distribution $\pi(\theta)$, is approximated by the empirical distribution of the samples $\hat{\pi}(\theta)$, and their corresponding normalized weights $\tilde{w}(\theta_{i})$'s:

\begin{equation*}\label{EQ:Empirical}
    \pi(\theta)\approx \hat{\pi}(\theta)=\sum_{i=1}^{N}\tilde{w}(\theta_{i})\delta(\theta_{i}),
\end{equation*}
where $\delta(\theta_{i})$ is the Dirac measure that puts unit mass at $\theta_{i}$ and $\sum_{i=1}^{N}\tilde{w}(\theta_{i})=1$.

Poor choices of importance distributions may yield inaccurate approximations of the target distribution \citep{doucet2000sequential} due to weight degeneracy and sample impoverishment. As a result, the bulk of the resampled particles, $\theta_{i}$, do not reside in the high-probability regions of $\pi(\theta)$. Weight degeneracy occurs when almost all of the samples drawn the importance function have near-zero importance weights leaving just a few samples with any significant weights. Multinomial resampling using the normalized importance weights $\tilde{w}(\theta_{i})$ can combat weight degeneracy by eliminating the particles with very small important weights and replicating those with higher weights \citep{gordon1993novel,doucet2000sequential}. After re-sampling, we reset all of the importance weights $w(\theta_{i})$ to $1/N$ and replace the weighted empirical distribution $\hat{\pi}(\theta)$ with an unweighted empirical distribution $\ddot{\pi}(\theta)$:

\begin{equation*}\label{EQ:Resample}
\ddot{\pi}(\theta)=\frac{1}{N}\sum_{i=1}^{N}N_{i}\delta(\theta_{i}),
\end{equation*}

where $N_{i}$ is the number of replicates corresponding to particle $\theta_{i}$ and $\sum_{i=1}^{N}N_{i}=N$.

Weight degeneracy can lead to sample impoverishment where a small subset of particles $\theta_{i}$'s are heavily replicated in the re-sampling step; hence,  few unique particle remain. The unweighted/re-sampled empirical distribution $\ddot{\pi}(\theta)$ may poorly approximate the true target distribution $\pi(\theta)$. To alleviate sample impoverishment, mixture approximations \citep{gordon1993novel} or kernel smoothing methods \citep{liu2001combined} can mutate or rejuvenate the replicated particles. However, these methods may not scale well to high-dimensional target distributions \citep{doucet2000sequential}.

An alternative method mutates the replicated particles with samples from $K(\theta_{i}^{(t-1)})$, the Metropolis-Hastings transition kernel \citep{gilks2001following}, whose stationary distribution is also the target distribution $\pi(\theta)$. Here we run $J$ Metropolis-Hastings updates for each particle $\theta_{i}$, for $i=1,...,N$. Other mutation schemes use genetic algorithms \citep{zhu2018new} or different transition kernels, $K(\cdot)$ \citep{papaioannou2016sequential,murray2016parallel}. The length of the Markov chain, $J$, will be short and dependent on computing resources. We set the $j$th sample drawn via MCMC as the mutated particle $\tilde{\theta}_{i}$. Since $\tilde{\theta}_{i} \sim \pi(\theta)$, the resulting empirical distribution $\tilde{\pi}(\theta)$ approximates the target distribution  $\pi(\theta)$:

\begin{equation*}\label{EQ:Mutation}
\pi(\theta)\approx\tilde{\pi}(\theta)=\sum_{i=1}^{N}\tilde{\theta}_{i}\delta(\tilde{\theta}_{i}).
\end{equation*}

Even with the mutation step, sampling-importance-resampling may incur large computational costs. Poor choices of importance distributions may result in extreme sample impoverishment, due to the large discrepancy between the importance and target distribution. Here, the mutation stage typically requires very long (and costly) chains of the Metropolis-Hastings algorithm to move the particles into the high-probability regions of the target distribution \citep{li2014fight}.  

\subsubsection*{Sequential sampling-importance-resampling}
Our fast particle-based approach addresses the limitations noted above. We propose a series of intermediate posterior distributions $\pi_{t}(\theta|Z)$, for $t=1,...,T$ which will act as importance and target distributions. Existing methods use intermediate posterior distributions for parameter estimation of static systems \citep{chopin2002sequential,papaioannou2016sequential,nguyen2014sequential}, uncertainty quantification for chemical processes \citep{kalyanaraman2016uncertainty}, and calculating maximum-likelihood estimates for hierarchical systems \citep{lele2007data}.

Intermediate posterior distributions can be generated using likelihood tempering \citep{chopin2002sequential,neal2001annealed, Liang_Wong_2001}. For each intermediate posterior distribution $\pi_{t}(\theta|Z)$, the likelihood component is a fractional power of the original likelihood $p(Z|\theta)$. The $t$th intermediate posterior distribution, $\pi_{t}(\theta)$, is generated as follows:

\begin{equation}\label{EQ:Incorporate}
\pi_{t}(\theta|Z) \propto  p(Z|\theta)^{\gamma_{t}} p(\theta),
\end{equation}
where $\gamma_t$'s are determined according to a schedule where $\gamma_{0} = 0 < \gamma_{1} < \cdots < \gamma_{T} = 1.$

For cycle $t=1$, we set the importance distribution to be the prior distribution $p(\theta)$, and the target distribution to be the first intermediate posterior distribution, $\pi_{1}(\theta|Z)$. For cycle $t$, the importance distribution is $\pi_{t-1}(\theta|Z)$ and the target distribution is $\pi_{t}(\theta|Z)$. Note that the likelihood incorporation schedule need not be uniform. For instance, more of the likelihood can be incorporated into the earlier intermediate posterior distributions. 

Finally, we mutate the particles via short runs of the Metropolis-Hastings algorithm, where the stationary distribution is $\pi_{t}(\theta)$, the $t$th intermediate posterior distribution. Note that the importance and target distributions are consecutive ($t$th and $t+1$th) intermediate posterior distributions, so there is considerable overlap between the high-probability regions of the two distributions. Convergence results for this family of Sequential Monte Carlo algorithms are provided in \citet{crisan2000convergence}, \citet{beskos2016convergence}, and \citet{giraud2017nonasymptotic}.

\subsection{Stopping criterion}
We present a stopping rule that controls the number of Metropolis-Hastings updates within the mutation step. This provides an automatic heuristic determining when to stop the mutation stage, and it can also eliminate unnecessary computer model runs. The stopping rule is based on the Bhattacharyya distance \citep{bhattacharyya1946measure}, $D_{B}(p,q)$, which measures the similarity between two distributions, $p(\theta)$ and $q(\theta)$. 
We first evaluate the stopping criterion after $2k$ Metropolis-Hastings updates; if the criterion is not met, then we re-evaluate after $k$ subsequent updates. 

Consider $\theta_{t}^{i,k}$, the $i$th particle, or parameter setting, after the $k$th mutation step of the Metropolis-Hastings algorithm during cycle number $t$. Let $\mathbf{\theta_{t}^{k}}=\{\theta_{t}^{1,k},...,\theta_{t}^{n,k}\}$ denote the set of parameters $\theta_{t}^{i,k}$'s. Let $h(\theta_{t}^{i,k})$ be the target metric of interest evaluated at parameter setting $\theta_{t}^{i,k}$, in this case, the Antarctic ice sheet contribution to sea level change in 2100. Let $\mathbf{h(\theta_{t}^{k})}=\{h(\theta_{t}^{1,k}),...,h(\theta_{t}^{n,k})\}$ denote the set of target metrics $h(\theta_{t}^{i,k})$'s. 

At mutation update $2k$, we partition the range spanned by two sets of target metric samples -- $\mathbf{h(\theta_{t}^{k})}$ and $\mathbf{h(\theta_{t}^{2k})}$ -- into $m$ non-overlapping blocks of equal width.  Then, we compute the real-valued Bhattacharyya distance $D_{B}(\mathbf{h(\theta_{t}^{k})},\mathbf{h(\theta_{t}^{2k})})=-\ln \Big(\sum _{{i=1}}^{n}{\sqrt  {p_{i}q_{i}}}\Big)$
where $p_{i}$ and $q_{i}$ are the proportion of samples, from $\mathbf{h(\theta_{t}^{k})}$ and $\mathbf{h(\theta_{t}^{2k})}$ respectively, that lie within the ith partition. The mutation step proceeds until $D_{B}(h(\theta_{k}),h(\theta_{2k})) < \epsilon_{B}$, the stopping criterion. If the stopping criterion is not fulfilled, we run $k$ additional Metropolis-Hastings updates and evaluate the stopping criterion at iterations $3k$ and $2k$. We repeat this until the stopping criterion is met. We obtain the threshold $\epsilon_{BD}$ through a Monte Carlo simulation run prior to the calibration study. Section \ref{Sec:Tuning} discusses tuning for $k$, $\epsilon_{B}$, and $m$. 

\subsection{Adaptive incorporation schedule}
In Equation \ref{EQ:Incorporate}, we introduce a standard incorporation schedule $\mathbf{\gamma}=(\gamma_0,...,\gamma_{T})$. In the standard implementation, the user must select the total number of sampling-importance-resampling cycles (T) and the likelihood incorporation increments $\gamma_t$ for $t=(0,...,T)$. Past research proposed novel methods to adaptive choose the incorporation schedule, $\gamma_t$, yet maintain a constant number of cycles, $T$ \citep{nguyen2014sequential,kalyanaraman2016uncertainty}. Here, we introduce an adaptive incorporation schedule that automatically determines both the total number of sampling-importance-resampling cycles, $T$, and incorporation schedule,$\mathbf{\gamma}$. Introducing the adaptive incorporation schedule into the particle-based calibration framework provides computational and practical benefits by (1) reducing the number of computer model evaluations; (2) decreasing the overall calibration wall times; and (3) simplifying implementation for the user. 

The adaptive incorporation schedule proceeds as follows. On initialization, we set the initial incorporation increment $\gamma_{0}$ to $0$. We draw the initial set of particles $\theta_{0}$ from $\pi_{0}(\theta|Z)\propto L(\theta|Z)^{0}p(\theta)=p(\theta)$, the prior distribution of model parameters. 
For cycle $t=1,2,3,...$, we calculate the full likelihood $L(\theta_{t-1}^{(i)}|Z)$ for $i=1,...,N$ where $\theta_{t-1}^{(i)}$ denotes the parameter samples from the previous cycle $t-1$. For computational efficiency, we reuse the likelihood evaluations from the previous cycle. Next, we find the optimal $\gamma_{t}$ that returns an effective sample size (ESS) of $ESS_{thresh}$ or a sample size closest to $ESS_{thresh}$: 
$\gamma_{t}= \argmin_{\gamma}\{(ESS_{\gamma}-ESS_{thresh})^{2}\}$
, where $\gamma\in (\gamma_{min},1-\gamma_{t-1})$, $\gamma_{min}$ is a previously set minimum incorporation value,  $ESS_{\gamma}=\sum_{i=1}^{N}\frac{1}{w_{t}^{(i){2}}}$, and $w_{t}^{(i)}\propto L(\theta_{t}^{(i)}|Z)^{\gamma}$. Note that we can lower computational costs by evaluating the full likelihood $L(\theta_{0}^{(i)}|Z)$ only once before the optimization. 

We stop the scheduling algorithm when $\sum_{i=1}^{t}\gamma_{t}=1$. This occurs when the entire likelihood has been incorporated, and the target distribution has evolved to the full posterior distribution $\pi(\theta|Z)$. Note at each cycle $t$, we set the incorporation increment ($\gamma_{t}$) to be between $\gamma_{min}$ and $1-\sum_{i=1}^{t}\gamma_{t}$. In Section \ref{Sec:Tuning}, we describe how to set the minimum incorporation increment $\gamma_{min}$ and the threshold effective sample size, $ESS_{thresh}$. 

\textbf{Adaptive likelihood incorporation schedule}

\begin{enumerate}
    \item Initialization: At $t=0$, set $\gamma_{0}=0$.
    \item When $t>0$ and $\sum_{i=1}^{t-1}\gamma_{i}<1$
    \begin{itemize}
        \item Compute $L(\theta_{t-1}^{(i)}|Z)$ for $i=1,...,N$
        \item Set $\gamma_{t}= \argmin_{\gamma}\{(ESS_{\gamma}-ESS_{thresh})^{2}\}$, where $ESS_{\gamma}=\sum_{i=1}^{N}\frac{1}{w_{t}^{(i){2}}}$,  $w_{t}^{(i)} \propto L(\theta_{t}^{(i)}|Z)^{\gamma}$, and $\gamma\in (\gamma_{min},1-\gamma_{t-1})$. 
        \item $\gamma_{min}$ is a predetermined minimum incorporation value
    \end{itemize}
    \item When $t>0$ and $\sum_{i=1}^{t-1}\gamma_{i}=1$: Stop Calibration
\end{enumerate}

\begin{algorithm}
 \KwData{ $Z$}
 \textbf{Initialization:} \\
 Draw $\theta_{0}^{(i)}\sim p(\theta)$ for particles $i=1,...,N$. \\
 Set $w^{(i)}_{0}=1/N$, $\gamma_{0}=0$, and $K$;
 
 \For{cycles $t=1,....,T$}{
  \textbf{1. Compute full likelihood:}\\
  Calculate $L(\theta_{t-1}^{(i)}|Z)$ for $i=1,...,N$;\\
  \textbf{2. Select optimal likelihood incorporation increment $\gamma_t$:}\\
  Set $\gamma_{t}= \argmin_{\gamma}\{(ESS_{\gamma_{t}}-ESS_{thresh})^{2}\}$, where $\gamma\in (0.1,1-\sum_{i=1}^{t-1}\gamma_{t-1})$\\
  Note: $ESS_{\gamma_{t}}=\sum_{i=1}^{N}\frac{1}{w_{t}^{(i){2}}}$ and $w_{t}^{(i)}\propto L(\theta_{t}^{(i)}|Z)^{\gamma_{t}}$;\\
  \textbf{3. Compute importance weights:}\\
  $w_{t}^{(i)}\propto w_{t-1}^{(i)}\times L(\theta_{t}^{(i)}|Z)^{\gamma_{t}}$;\\
  \textbf{4. Re-sample particles:}\\
  Draw $\theta_{t}^{(i)}$ from $\{\theta_{t-1}^{(1)},...,\theta_{t-1}^{(N)}\}$ with probabilities $\propto \{w_{t}^{(1)},...,w_{t}^{(N)}\}$;\\
  \textbf{5. Set intermediate posterior distribution:}\\
  Set $\pi_{t}(\theta|Z)\propto L(\theta_{i}|Z)^{\tilde{\gamma}}\pi(\theta)$, where $\tilde{\gamma}=\sum_{j=1}^{t}\gamma_{j}$;\\
  \textbf{6. Mutation:} \\
  Using each particle $(\theta_{t}^{(1)},...,\theta_{t}^{(N)})$ as the initial value, run $N$ chains of an MCMC algorithm with target distribution $\pi_{t}(\theta|Z)$ for $2K$ iterations\\
  \textbf{7. Check stopping criterion:} \\
  Compute $\delta_{B}=D_{B}(h(\theta_{t}^{K}),h(\theta_{t}^{2K}))$;\\
  \eIf{$\delta_{B}<\epsilon_{B}$}{
   Set $\theta_{t}^{(i)}= \theta_{t}^{(i),2K}$;
   }{
   Run $K$ additional updates and re-evaluate stopping criterion\\
   Continue until stopping criterion is met
  }
\textbf{8. Stop when full likelihood is incorporated}\;
  \eIf{$\sum_{i=1}^{N}\gamma_{t}=1$ }{
   End Algorithm\;
   }{
   \textbf{Reset weights:}  $w_{t}^{(i)}=1/N$ for particles $i=1,...,N$\;
   Set t=t+1 and return to Step 1\;
  }
 }
 \caption{Fast Particle-based Calibration}
\end{algorithm}

\subsection{Tuning the algorithm}\label{Sec:Tuning}
Much of the algorithm above is automated. However, the user needs to choose: (1) the total number of particles, $N$; (2) the number of Metropolis-Hastings updates run before checking the stopping criterion, $K$; (3) the minimum incorporation $\gamma_{min}$; and (4) the effective sample size threshold $ESS_{thresh}$. (1) and (2) should be set based on the amount of available computational resources, but our simulation study results favor having more particles $N$ than longer Metropolis-Hastings updates $K$. We chose 2015 particles, which requires 56 nodes with 36 processors per node; thereby leaving one processor to execute master tasks. We set the reference length $k$ for the Metropolis-Hastings updates to be 7. Based on simulation experiments, the empirical distribution of particles stabilize after $10$ to $15$ updates. In this study, we set the floor for the incorporation increment, $\gamma_{min}$ to be 0.1 so that at each cycle, the weights for the importance sampling step is at least $L(\theta|Z)^{0.1}$. We set the threshold for the effective sample size  $ESS_{thresh}=N/2$.

We obtain $\epsilon_{BD}$ as follows. Prior to  running the calibration algorithm, we obtained samples of a target metric (Antarctic Ice Sheet contribution to sea level rise in 2100) from an initial survey of computer model runs. Let $\mu$ and $\sigma^2$ denote the sample mean and variance of the target metric mentioned above. We generate a collection of $B$ samples of size $n$, denoted as $\mathbf{x}=\{x_{1},...,x_{B}\}$. Here, $x_{b}\sim\mathcal{N}(\mu, \sigma^2)$, with $\mu$ and $\sigma^2$ previously defined.  Let $x_{base}\sim \mathcal{N}(\mu, \sigma^2)$ be a baseline sample for calculating the Bhattacharrya distance. We calculate $D_{B}(x_{b},x_{base})$ for $b=1,...,B$, and set $\epsilon_{BD}$ to be the $0.975$ quantile. In this study, we chose $B=1000$ and the number of partitions $m=200$. 

We calibrate the PSU3D-ICE model using Cheyenne \citep{cheyenne2017}, a 5.34-petaflops high performance computer operated by the National Center for Atmospheric Research (NCAR). Parallelized operations, such as calculating importance weights and mutation, proceed via message passing interface (MPI). To limit communication costs, we build the ice sheet model and load the relevant datasets separately on each processor. 

\subsection{Computational advantages and limitations}
We take advantage of the embarrassingly parallel nature of the importance sampling and mutation steps to reduce wall time. In our approach, the Metropolis-Hastings updates in the mutation stage are the primary drivers of computational cost. To address this cost, we propose an automated stopping rule for the mutation stage. We also introduce an adaptive likelihood incorporation schedule that automatically selects an efficient number of sampling-importance-resampling cycles. The stopping rule and adaptive likelihood incorporation schedule simplifies implementation for the user (due to automation) and reduces the number of computer model runs needed for calibration.

Our approach is a viable alternative to existing calibration methods, which may be computationally infeasible. MCMC-based calibration methods using the computer model is computationally prohibitive due to the sequential nature of MCMC algorithms.  Emulation-calibration methods, while efficient for expensive computer models, do not easily scale to problems with many parameters (say more than five or six for this model). 
Also, multiple-try MCMC methods \citep{liu2000multiple}, a mixture of importance sampling and MCMC, may incur large costs because several parallel processes must be initialized and terminated at each iteration of the MCMC chain. 
Multiple-try MCMC may experience slow mixing, especially when the Markov chain moves to the low-probability regions of the target distribution distribution \citep{martino2018review}.

While our method has many computationally advantages, we note that the heavy parallelization in our approach requires access to and the ability to work with high performance computing resources. Given our current computing resources, our method is ideally suited to models that run between six seconds and 15 minutes. For models with longer run times, the computational costs remain prohibitive. MCMC algorithms may be feasible and simpler to implement for models with shorter run times. As is the case with parallel computing methods, communication costs must be small relative to the computer model run times; otherwise we would not reap the benefits of our approach. 

\section{Simulated example and results}\label{Sec:Simulation}
In this section, we calibrate a simple computer model using three different methods. We simulate a data set of size $n=300$ where the spatial locations $s_{i}$ for $i=1,...,n$ are in the unit domain $[0, 1]^2$. We generate the data via a modified version of the computer model presented in \citet{bayarri2007computer}. We construct a simple computer model as follows:
\begin{equation*}
\label{eq:toymodel}
Y(s_{i},\theta)=5\times\exp \{-\theta(\mbox{lat}_{i} \times \mbox{lon}_{i}) \},
\end{equation*}
where $Y(s_{i},\theta)$ is a real-valued computer model output at model parameter setting $\theta$ and at a spatial location specified by lat$_{i}$ and lon$_{i}$, which represent the latitude and longitude of the $i$th location. The true process includes a data-model discrepancy term $\delta(s_{i})$ , which is defined as $\delta(s_{i})=-1.5\times(\mbox{lat}_{i} \times \mbox{lon}_{i})$, and iid observational error $\epsilon_{i}\sim \mathcal{N}(0,\sigma^{2}_{\epsilon})$. For this example, we set $\theta=1.7$ and $\sigma^{2}_{\epsilon}=0.5$. To generate the observational data, $Z(s_{i})$, we combine the computer model output $Y(s_{i},\theta)$, the data-model discrepancy, $\delta(s_{i})$, and the observational error, $\epsilon_{i}$, as follows:

$$Z(s_{i})= Y(s_{i},\theta)+\delta(s_{i})+\epsilon_{i}.$$

We model the observations as

$$Z(s_{i})=5\times\exp \{-\theta(\mbox{lat}_{i} \times \mbox{lon}_{i}) \}+\delta(s_{i})+\epsilon_{i},$$

where $\epsilon_{i} \sim \mathcal{N}(0,\sigma^{2}_{\epsilon})$ are the iid observational errors. Since the actual form of the discrepancy term is unknown, we model the discrepancy $\delta(s_{i})$, as a zero-mean Gaussian process, $\delta(s)\sim \mathcal{GP}(0,\Sigma_{\delta}(\xi_{\delta}))$, where $\xi_{\delta}$ is a vector containing the covariance parameters. To allow for some roughness of the process between spatial locations we choose an exponential covariance function $\Sigma_{\delta}(\xi_{\delta})= \sigma^{2}_{\delta}\exp\left(-\frac{|s_{i}-s_{j}|}{\phi_{\delta}}\right)$ with $\xi_{\delta}=(\phi_{\delta},\sigma^{2}_{\delta})$. To complete the Bayesian framework, we use the prior distributions: $\theta \sim \mathcal{N}(0,100)$, $\sigma^{2}_{\epsilon} \sim \mathcal{IG}(2,2)$, $\phi_{\delta} \sim \mathcal{U}(0.01,1.5)$, and $\sigma^{2}_{\delta} \sim \mathcal{IG}(2,2)$.

 We compare results from three calibration methods: (1) MCMC-based, (2)  standard particle-based, and (3) adaptive particle-based. In the MCMC-based method, we generated $100,000$ samples from $\pi(\theta,\phi_{\delta},\sigma^{2}_{\delta},\sigma^{2}_{\epsilon}|Z)$ via the Metropolis-Hastings algorithm. Next, the standard and adaptive particle-based calibration methods use $N=2000$ particles to approximate $\pi(\theta,\phi_{\delta},\sigma^{2}_{\delta},\sigma^{2}_{\epsilon}|Z)$. For the standard particle-based method, we set the total number of cycles to be 10, and establish a uniform likelihood incorporation $\mathbf{\gamma}=(\gamma_{1},...,\gamma_{10})$, where $\gamma_{t}=0.1$ for $t=1,...,10$. We run $K=100$ Metropolis-Hastings updates for each mutation cycle. In the adaptive particle-based calibration approach, our algorithm automatically chose four cycles with a likelihood incorporation schedule $\mathbf{\gamma}=(\gamma_{1},\gamma_{2},\gamma_{3},\gamma_{4})=(0.100, 0.148, 0.2743, 0.4777)$ using the adaptive likelihood incorporation schedule. For each mutation step, we run batches of $K=5$ Metropolis-Hastings updates until the stopping criterion is met. 

All three methods yield comparable calibration results (see Table \ref{Table:Toy}); however, our adaptive particle-based approach exhibits a considerable speedup in computation. For the model parameter, $\theta$, calibration via  MCMC (the "gold standard") provides estimate $\hat{\theta}_{mcmc}=2.04$ and $95\%$ credible interval bounds $(1.06, 3.17)$. Similarly, the standard particle-based approach generates estimate $\hat{\theta}_{std}=2.04$ with $95\%$ credible interval bounds $(1.03, 3.11)$ and the adaptive particle-based approach yields estimate $\hat{\theta}_{adapt}=2.04$ with $95\%$ credible interval bounds $(1.05, 3.14)$. The adaptive particle-based approach has  considerably shorter wall times due to fewer computer model evaluations. To illustrate, the adaptive approach requires just $10\times4=40$ sequential computer model runs, as opposed to $10\times100=1000$ runs for the standard particle-based approach and $100,000$ for the MCMC-based approach. 

\begin{table}[h]
\caption{Simulated example calibration results for three calibration approaches: (1) Adaptive particle-based; (2) Standard Particle-based; and (3) MCMC-based. All three approaches yield comparative results.}
\centering
\begin{tabular}{rllll}
  \hline
 & $\theta$ & $\phi_{\delta}$ & $\sigma^{2}_{\delta}$ & $\sigma^{2}_{\epsilon}$ \\ 
  \hline
  Adaptive Particle-Based (Est) & 2.04 & 1.22 & 0.78 & 0.44 \\ 
  Adaptive Particle-Based (95\% CI) & (1.05,3.14) & (0.83,1.50) & (0.36,1.32) & (0.36,0.52) \\ 
  Standard Particle-Based (Est) & 2.04 & 1.22 & 0.80 & 0.44 \\ 
  Standard Particle-Based (95\% CI) & (1.03,3.11) & (0.81,1.50) & (0.32,1.33) & (0.35,0.51) \\ 
  MCMC-Based (Est) & 2.04 & 1.21 & 0.79 & 0.44 \\ 
  MCMC-Based (95\% CI) & (1.06,3.17) & (0.80,1.50) & (0.34,1.33) & (0.36,0.52) \\ 
   \hline
\end{tabular}

\label{Table:Toy}
\end{table}

\section{Application to the PSU3D-ICE model}\label{Sec:Results}
Here we provide specifics for calibrating the PSU3D-ICE model and discuss how our method provides key computational benefits over existing calibration approaches. We also summarize results from a comparative analysis of three calibration methods within the context of the PSU3D-ICE model. The efficiency of our computational approach allows us to study the effect of observations from the Pliocene on parameter calibration and projections of sea level rise and also enables us to conduct a prior sensitivity analysis.

\subsection{Calibrating PSU3D-ICE}
We calibrate 11 model parameters using both paleoclimate records and modern observations from satellite imagery (Section \ref{Sec:PSU3DICE}). For the paleoclimate records, modern volume, and modern grounded ice area, we use independent truncated normal distributions. The upper and lower ranges for the truncated normal likelihood functions are based on domain area expertise and past studies (Section \ref{Sec:Observations}).

We calibrate the PSU3D-ICE model using five observations: (1) $Z_{plio}$, the Antarctic ice sheet's contribution to sea level change ($m$) in the Pliocene; (2) $Z_{lig}$, contribution in the Last Interglacial Age ($m$); (3) $Z_{lgm}$, contribution in the Last Glacial Maximum ($m$); (4) $Z_{vol}$, the Antarctic ice sheet's total ice volume in the modern era ($km^{3}$) ; and (5) $Z_{area}$, total grounded ice area in the modern era ($km^{2}$). We also use observations of ice occurrence taken at 10 strategic point in the Antarctic Ice Sheet. Here, $Z_{spat}= (Z_{spat,1},...,Z_{spat,10})$. All ten locations have ice presence; so, $Z_{spat,i}=1$ for locations $i=1,...,10$. 

\subsubsection*{Likelihood} \label{Sec:Likelihood}
For the observational dataset, $Z=(Z_{plio}, Z_{lig}, Z_{lgm}, Z_{vol}, Z_{area},Z_{spat,1},..., Z_{spat,10})$, we define a likelihood function using truncated normal distributions and indicator functions. For the modern volume, modern total grounded area, and paleoclimate records, we use independent truncated normal distributions as the observational model. $TN(\mu,\sigma^{2},\alpha, \beta)$ denotes a truncated normal distributions with the mean ($\mu$), variance ($\sigma^2$), upper bound ($\alpha$), and lower bound ($\beta$).

$$Z_{plio}\sim \mathcal{TN}(\mu=Y(\theta)_{plio}, \sigma^{2}=30^2,\alpha=Y(\theta)_{plio} - 10, \beta=Y(\theta)_{plio} + 10)$$
$$Z_{lig}\sim \mathcal{TN}(\mu=Y(\theta)_{lig}, \sigma^{2}=10^2,\alpha=Y(\theta)_{lig} - 2, \beta=Y(\theta)_{lig} + 2)$$
$$Z_{lgm}\sim \mathcal{TN}(\mu=Y(\theta)_{lgm}, \sigma^{2}=20^2,\alpha=Y(\theta)_{lgm} - 5, \beta=Y(\theta)_{lig} + 5)$$
$$Z_{vol}\sim \mathcal{TN}(\mu=Y(\theta)_{vol},\sigma^{2}=1.6\times 10^{15},\alpha=Y(\theta)_{vol}- 2.5\times 10^{15},\beta=Y(\theta)_{vol}+ 2.5\times 10^{15})$$
$$Z_{area}\sim \mathcal{TN}(\mu=Y(\theta)_{area},\sigma^{2}=0.6\times 10^{12},\alpha=Y(\theta)_{area}- 1.5\times 10^{12},\beta=Y(\theta)_{area}+ 1.5\times 10^{12})$$

The second set of observations are binary occurrences of ice at 10 strategically placed points on the Antarctic ice sheet (Supplement).  For these observations, we use indicator functions as the observational model as follows:
$$Z_{spat}\sim \prod_{i=1}^{10}\mathbb{I}(Y(\theta)_{spat,i}=Z_{spat,i}),$$
where $Y(\theta)_{spat,i}$ denotes the model spatial output for a model run using parameters $\theta$. 

\subsubsection*{Priors}\label{Sec:Priors}
We set the prior distributions for the 11 model parameters based on expert knowledge. Five model parameters - CALVNICK, TAUASTH, CALVLIQ, CLIFFVMAX, FACEMELTRATE - have uniform prior distributions. Here, $\theta\sim U(\alpha,\beta)$, where $\alpha$ and $\beta$ denote the upper and lower bounds of the uniform distribution. The prior distributions are as follows: 
\begin{multicols}{2}
\begin{itemize}
    \item $\theta_{CALVNICK}\sim \mathcal{U}(0,2)$
    \item $\theta_{TAUASTH}\sim \mathcal{U}(1000,5000)$
    \item $\theta_{CALVLIQ}\sim \mathcal{U}(0,200)$
    \item $\theta_{CLIFFVMAX}\sim \mathcal{U}(0,12000)$
    \item $\theta_{FACEMELTRATE}\sim \mathcal{U}(0,20)$
\end{itemize}
\end{multicols}

Six parameters - OCFACMULT, OCFACMULTASE, CRHSHELF, ENHANCESHEET, ENHANCESHELF, CRHFAC - have log-uniform prior distributions. Here, $\theta\sim LU(base,\alpha,\beta)$, which implies $log_{base}(\theta)\sim U(\alpha,\beta)$ where $\alpha$ and $\beta$ denote the upper and lower bounds of the uniform distribution. The prior distributions are as follows: 
\begin{multicols}{2}
\begin{itemize}
\item $\log_{10}(\theta_{OCFACMULT})\sim \mathcal{U}(-0.5,0.5)$
\item $\log_{10}(\theta_{OCFACMULTASE})\sim \mathcal{U}(0,1)$
\item $\log_{10}(\theta_{CRHSHELF})\sim \mathcal{U}(-7,-4)$
\item $\log_{10}(\theta_{ENHANCESHEET})\sim \mathcal{U}(-1,1)$
\item $\log_{0.3}(\theta_{ENHANCESHELF})\sim \mathcal{U}(-1,1)$
\item $\log_{10}(\theta_{CRHFAC})\sim \mathcal{U}(-2,2)$
\end{itemize}
\end{multicols}

We can estimate the data-model discrepancy as an additive model bias, $\alpha\in \mathbb{R}$, such that our observational model (\ref{EQ:ModelCalibration}) is modified to be $Z=Y(\theta)+\alpha+\epsilon$. For observations that are discontinuous in time, past ice sheet calibration studies \citep{Edwards2019, williamson2013history,ruckert2017assessing} model the discrepancy term as a tolerance to the observation measurement error, which follows the zero-mean Gaussian process framework provided in \citet{kennedy2001bayesian}. For the PSU3D-ICE model, we find that calibration with and without the discrepancy term yields very similar results. 

\subsection{Computational benefits of our approach}

\begin{figure}[ht]
    \centering
    \includegraphics[width=1\textwidth]{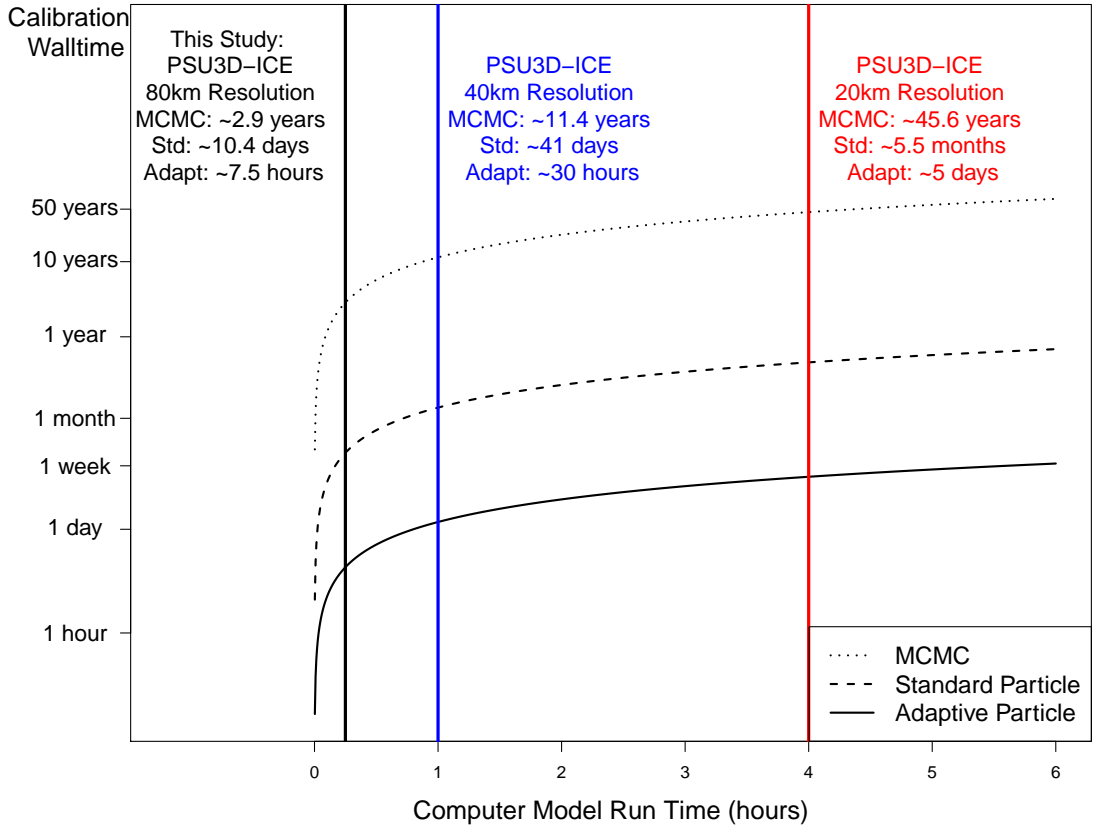}
    \caption{Total calibration wall time (y-axis) with respect to length of single computer model run (x-axis). Dotted lines represent calibration using traditional Markov Chain Monte Carlo (MCMC). Dashed lines represent calibration using a standard particle-based approach. Solid lines denote calibration via an adaptive particle-based approach (our approach). Vertical colored lines denote the model run times for the PSU3D-ICE model at different spatial resolutions. The adaptive particle-based approach shows a dramatic speed-up over traditional MCMC-based methods.}
    \label{Fig:Cost}
\end{figure}

Our adaptive particle-based approach greatly reduces calibration wall times compared to other calibration methods. Figure \ref{Fig:Cost} provides approximate calibration wall times for the PSU3D-ICE model across three calibration methods: (1) MCMC-based; (2) standard particle-based; and (3) adaptive particle-based. Figure \ref{Fig:Cost} also illustrates how wall times increase with model complexity. 

The computing time for our approach is based on the time taken to run the PSU3D-ICE model at $80$ km spatial resolution and an adaptive temporal resolution with a baseline timestep of 8 years. Run times are for the NCAR Cheyenne HPC system with 2.3-GHz Intel Xeon E5-2697V4 Broadwell processors. To approximate the wall times as a function of computer model complexity, we scale the wall time according to its cost relative to the baseline PSU3D-ICE model above. 
Note that in practice, computation times for the particle-based methods can be slightly higher due to initialization and communications costs inherent to parallelized computing. Reduction of initialization and communication costs is an active area of research with novel methods in development \citep{ballard2016reducing,fan2018parallelizing}. The calibration time for the MCMC approach is the estimated time to generate 100k samples using a simple all-at-once random-walk Metropolis-Hastings algorithm. Calibration times for the standard particle-based approach is based on using $N=2015$ particles, $T=10$ importance sampling cycles, and $K=100$ Metropolis-Hastings updates in the mutation stage.

\subsection{Comparisons to other calibration approaches}
We conduct a comparative study between our particle-based calibration approach and competing emulation-calibration methods (see Appendix for details). We calibrate the PSU3D-ICE model using three methods:
\begin{enumerate}
    \item A low-dimensional emulation-calibration approach: This approach varies only three parameters -- OCFACMULT, CALVLIQ and CLIFFVMAX -- and fixes the remaining eight parameters at scientifically justified values provided by our expert on ice sheets (DP). We include this approach because reducing the number of parameters is a common way to address computational challenges associated with calibration with long model run times \citep[e.g.][]{Edwards2019,chang2014fast,sacks1989design}. We chose these three parameters because they are considered to be important in modeling the long-term evolution of the Antarctic ice sheet \citep{Edwards2019, deconto2016contribution}. We train a Gaussian process emulator using 512 design points and use the squared exponential covariance function to represent the dependence between the design points. For the experimental design, we use a full factorial design with eight equally spaced points for each model parameter.
    \item A high-dimensional emulation-calibration approach: This approach calibrates all 11 selected parameters of the PSU3D-ICE model. We train a Gaussian process emulator using 512 design points generated via Latin Hypercube Design (LHC). Similar to the low-dimensional case, we use an exponential covariance function to model the dependence between design points. Emulation and calibration details are provided in the Appendix.
    \item Our particle-based approach: We use our heavily parallelized particle-based approach to calibrate all 11 selected parameters. 
\end{enumerate}

\begin{figure}[ht]
    \centering
    \includegraphics[width=1\textwidth]{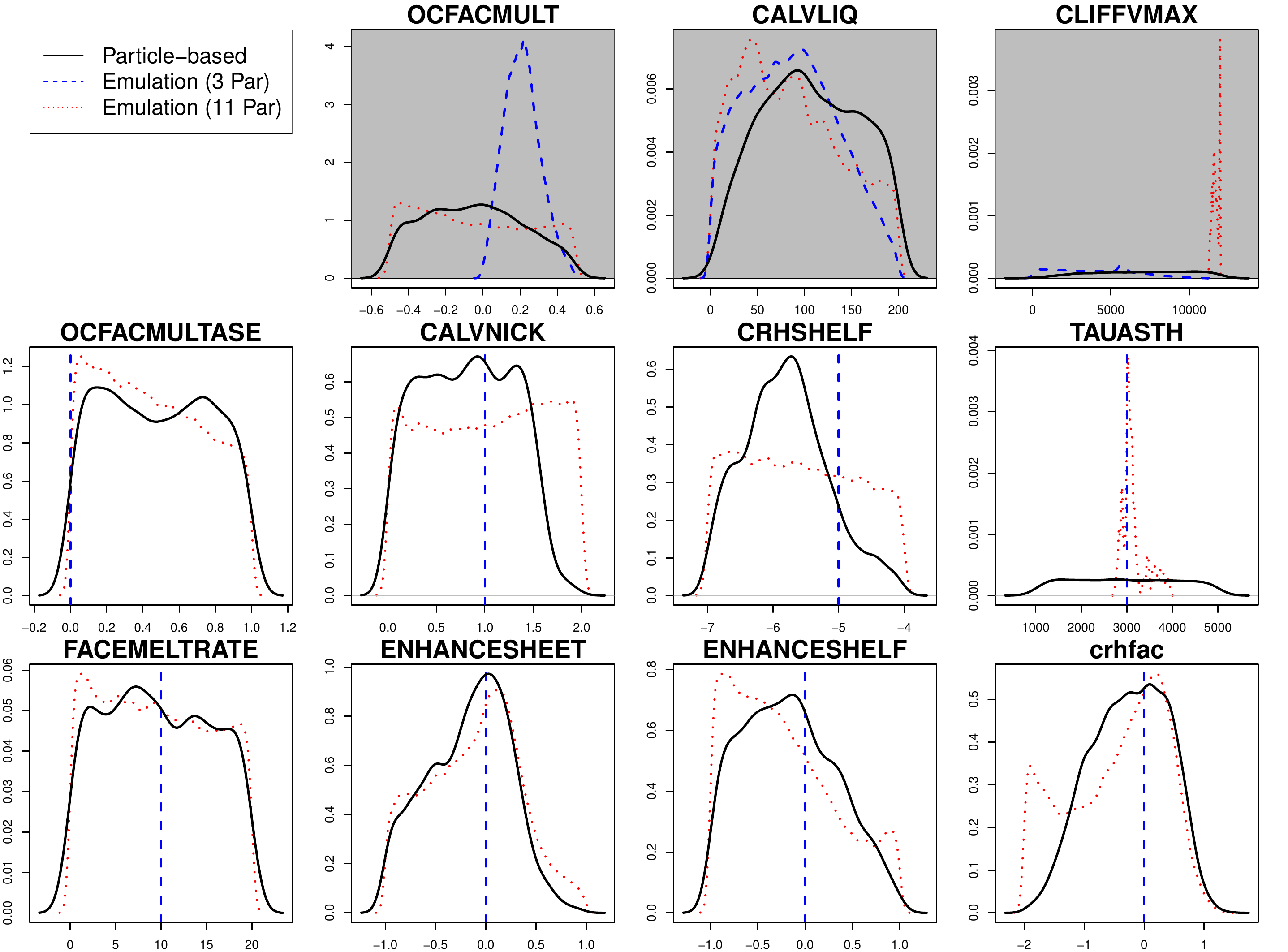}
    \caption{Posterior densities of model parameters using the adaptive particle-based approach (solid black line), emulation calibration with three parameters (dashed blue line), and emulation calibration with 11 parameters (dotted red line). Three-parameter emulation-calibration experiment use model parameters OCFACMULT, CALVLIQ, and CLIFFVMAX. The 11-parameter emulation-calibration experiment include all model parameters. Shaded panels denote parameters used in the three-parameter emulation-calibration experiment. Both emulation-based approaches result in sharper posterior densities than the particle-based approach for a subset of the model parameters.}
    \label{Fig:EmulPar}
\end{figure}

\begin{figure}[ht]
    \centering
    \includegraphics[width=1\textwidth]{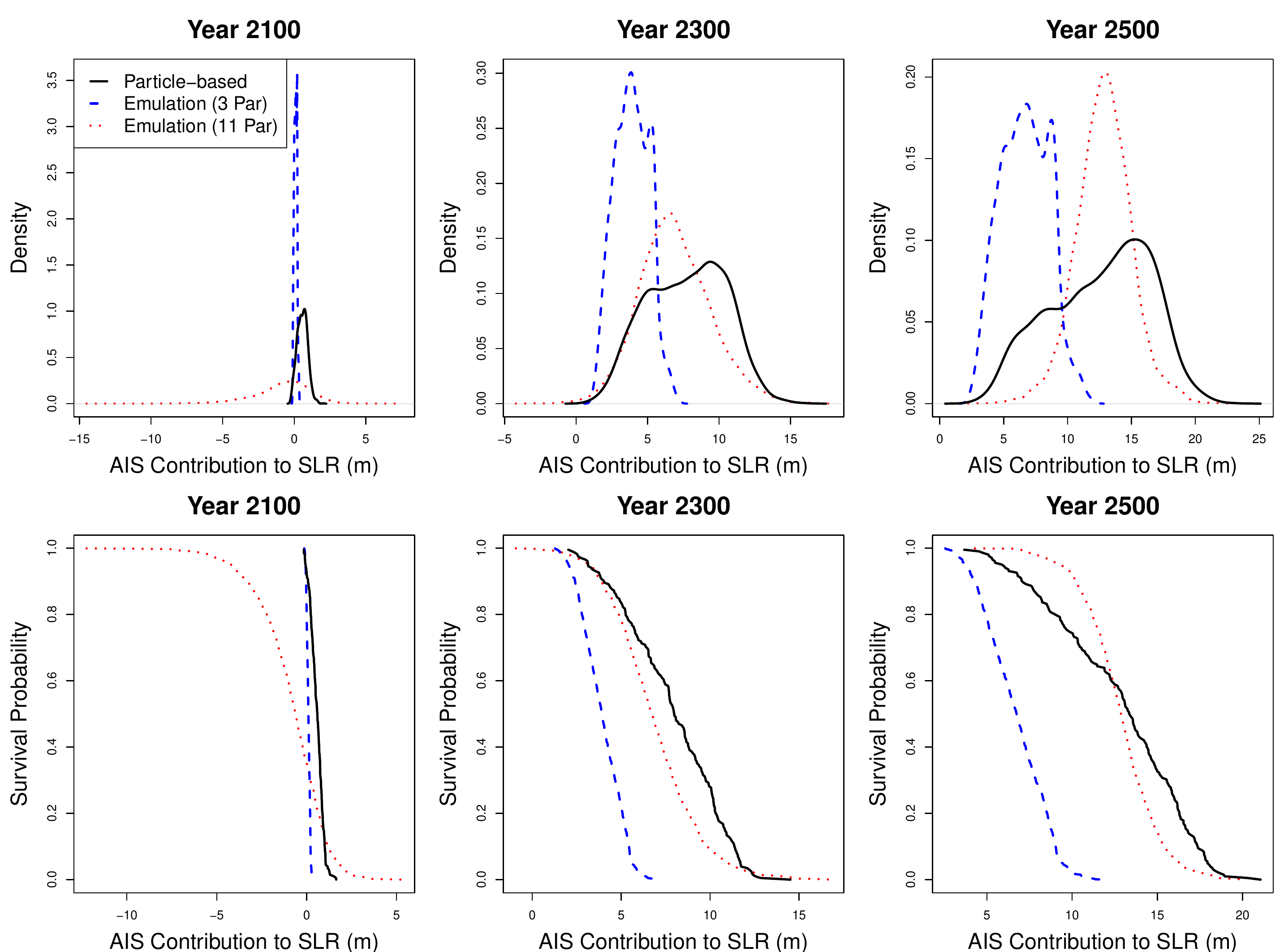}
    \caption{(Top Panel) Posterior densities of the projected Antarctic ice sheet's contribution to sea level change in 2100, 2300, and 2500 using the adaptive particle-based approach (solid black line), emulation calibration with three parameters (dashed blue line), and emulation calibration with 11 parameters (dotted red line). (Bottom Panel) Empirical survival functions of the projected Antarctic ice sheet's contribution to sea level change in 2100, 2300, and 2500 using the adaptive particle-based approach (solid black line), emulation calibration with three parameters (dashed blue line), and emulation calibration with 11 parameters (dotted red line). Three-parameter emulation results in sharper densities centered on distinctively lower point estimates. The 11-parameter emulation-calibration approach results in highly uncertain projections.}
    \label{Fig:EmulProj}
\end{figure}

For the first method, we find that by fixing eight of eleven parameters, we greatly constrain the parameter space and thereby underestimate the parametric uncertainty underlying the ice sheet model. Projections for the Antarctic sea level contribution in 2100-2500 are much lower and overconfident compared to those from our particle-based approach (Figure \ref{Fig:EmulProj}). For the second method, the limited amount of design points (training data) generates an inaccurate surrogate model as shown by the large out-of-sample cross-validated mean squared prediction error (Appendix). This calls into question the parameter estimates as well as the resulting projections. As shown in Figure \ref{Fig:EmulPar}, the second approach produces extremely sharp posterior distributions for two key model parameters, CLIFFVMAX and TAUASTH, which is inconsistent with the parameter estimates from the particle-based approach. 

\begin{figure}[h]
    \centering
    \includegraphics[width=0.9\textwidth]{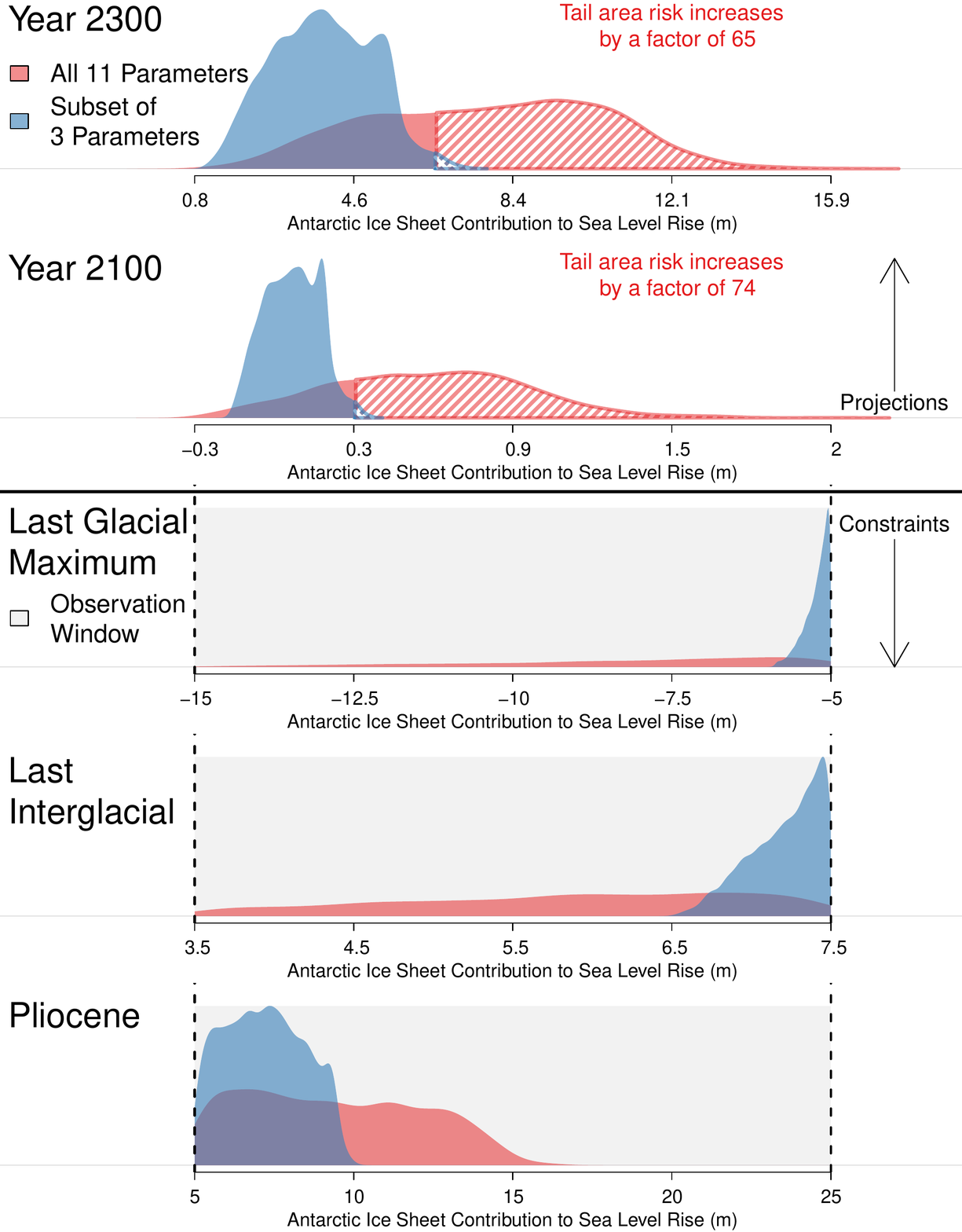}
    \caption{Antarctic ice Sheet contribution to sea level rise in the Pliocene (bottom panel), Last Interglacial Age (fourth panel), Last Glacial Maximum (third panel), 2100 (second panel), and 2300 (first panel). Red shading denotes the posterior densities for each time period and projections after calibrating 11 parameters using our fast particle-based approach. Blue shading denotes the posterior densities after calibrating three parameters using emulation-calibration. The light gray shading represents the observational constraints for the Last Glacial Maximum, Last Interglacial Age, and Pliocene. The striped red and striped blue shading represents the 99\%th percent quantile for the 11-parameter approach and three-parameter approach, respectively.}
    \label{Fig:TailAreaRisk}
\end{figure}

Figure \ref{Fig:TailAreaRisk} compares the posterior densities of projections and hindcasts for the three-parameter emulation-calibration approach and our 11-parameter particle-based method. Note that the three-parameter emulation-calibration approach (striped blue shading) underestimates the tail-area risk, or the 99-th\% quantile, for sea level projections compared to our approach (striped red shading). By calibrating more parameters, we can expect the tail-area risk to increase by a factor of 74 in 2100 and 65 in 2300.
    
The three-parameter emulation calibration required 1.5 minutes to fit the Gaussian process emulator using 12 processors on the Cheyenne HPC system and $\sim$1.5 hours to generate 500k samples via MCMC from the posterior distribution. The 11-parameter emulation calibration required 10 minutes to fit the emulator using 12 processors on the Cheyenne HPC system and $\sim$1.5 hours to generate 500k samples via MCMC from the posterior distribution.

\subsection{The effect of deep time observations on projections} 

Calibration can be improved by considering an important source of uncertainty, the state of the Antarctic ice sheet during the Pliocene era \citep{dolan2018high,Salzmann2013, dutton2015sea}. There is some evidence that the Antarctic ice sheet experienced fluctuations in volume during the Pliocene era \citep{naish2009obliquity}. Other studies suggest that at peak warming episodes during the Pliocene era, the Antarctic ice sheet had a lower volume, contributing to higher sea level rise \citep{cook2014sea,dolan2011sensitivity,dowsett1990high,pollard2009modelling, pollard2015potential,de2014fully}. However, the maximum Antarctic ice retreat and sea level rise contribution during the Pliocene remains largely uncertain \citep{dutton2015sea, rovere2014mid}.

\begin{figure}[ht]
    \centering
    \includegraphics[width=1\textwidth]{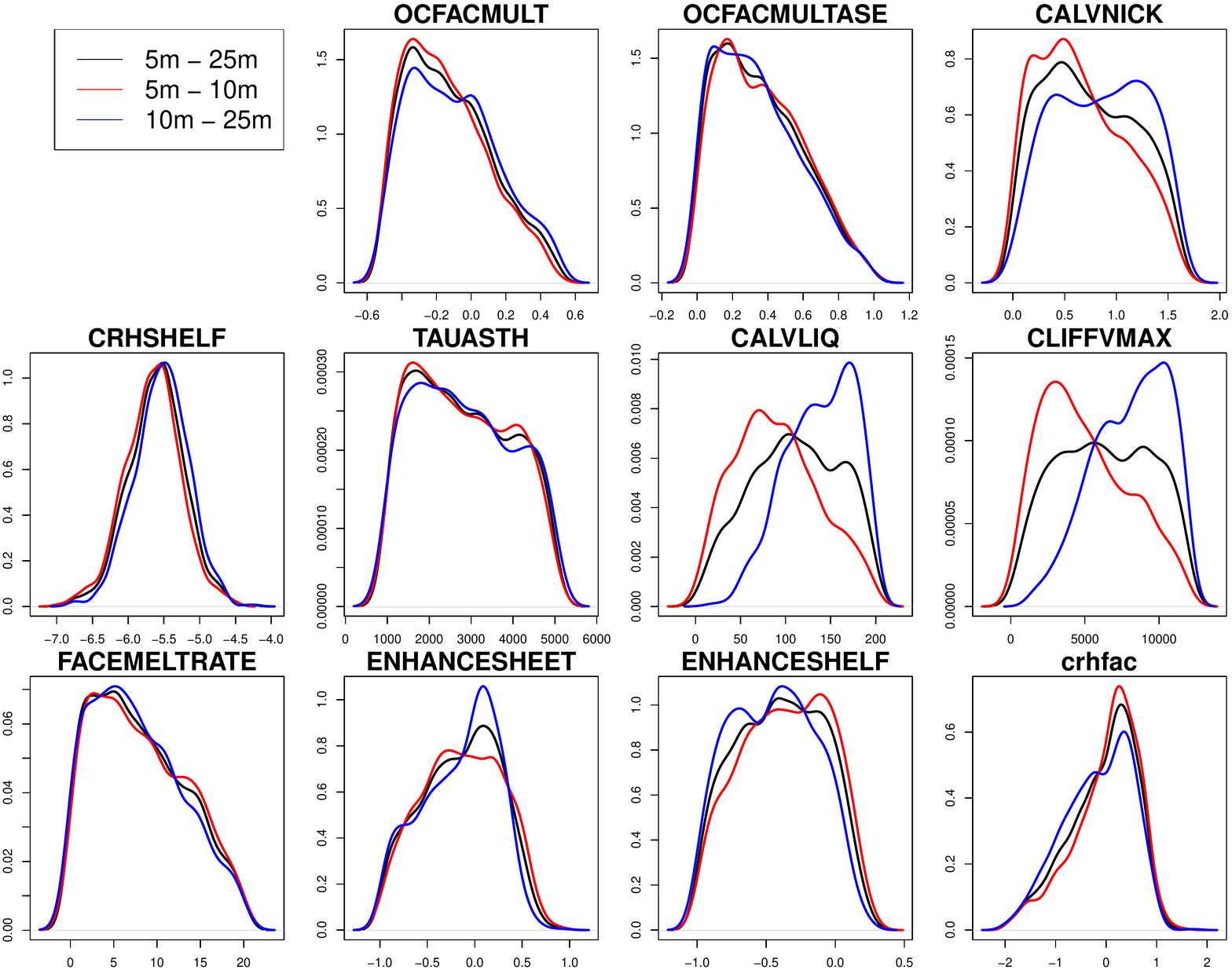}
    \caption{Posterior densities of model parameters for calibration using a wide Pliocene window of 5 m to 25 m (black line), low window of 5 m to 10 m (red line), and a high window of 10 m to 25 m (blue line). There is noticeable change in the densities for three model parameters - CALVNICK, CALVLIQ, and CLIFFVMAX.}
    \label{Fig:PlioPar}
\end{figure}

\begin{figure}[ht]
    \centering
    \includegraphics[width=1\textwidth]{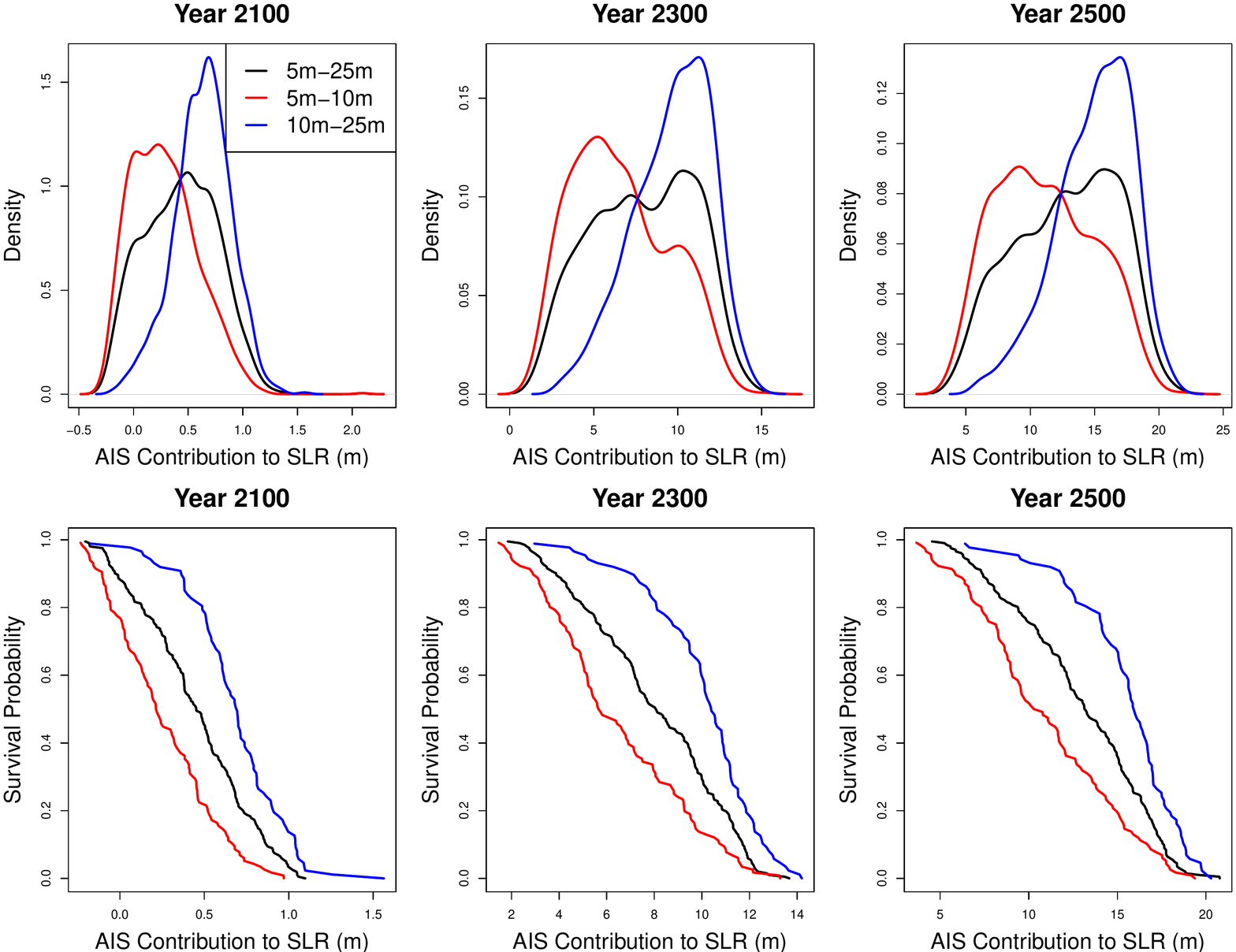}
     \caption{(Top Panel) Posterior densities of the projected Antarctic ice sheet's contribution to sea level change in 2100, 2300, and 2500 for calibration using a wide Pliocene window of 5 m to 25 m (black line), low window of 5 m to 10 m (red line), and a high window of 10 m to 25 m (blue line). (Bottom Panel) Empirical survival function of the projected Antarctic ice sheet's contribution to sea level change in 2100, 2300, and 2500 for calibration using a wide Pliocene window of 5 m to 25 m (black line), low window of 5 m to 10 m (red line), and a high window of 10 m to 25 m (blue line). constraining the Pliocene windows yield sharper projections of sea level rise. The higher window results in considerably higher projections than the lower window. }
    \label{Fig:PlioProj}
\end{figure}

We examine whether the width of the Pliocene observation windows (5 m to 25 m, 5 m to 10 m, 10 m to 25 m) has an influence on sea level projections and parameter estimation. (See Appendix for details on how these windows affect the likelihood function.)  
  Our results demonstrate that information regarding the nature of the Antarctic ice sheet during the Pliocene era has a strong influence on sea level projections. Figure \ref{Fig:PlioPar} illustrates how the posterior densities for two key model parameters (CALVLIQ and CLIFFVMAX) differ under the three Pliocene windows. Both parameters influence ice dynamics inherent to marine cliff instability (MICI) -- hydrofracturing due to surface melt (CALVLIQ) and structural failure of tall ice cliffs (CLIFFVMAX). As shown in Figure \ref{Fig:PlioProj}, increasing the Pliocene window from the range 5 m to 10 m to the range 10 m to 25 m requires more aggressive MICI (larger values of these parameters); hence resulting in higher projections of sea level rise (e.g. exceeding 3 m in 2300). If we are very uncertain about the Pliocene (represented by a very large window of 5 m to 25 m), the resulting sea level projections in 2300 also become highly uncertain (95\% credible interval of 1.2 m to 12.4 m), compared to projections from narrower windows of 5 m to 10 m (95\% credible interval of 1.2 m to 11.5 m) or 10 m to 25 m (95\% credible interval of 3.0 m to 12.9 m). 
  
\subsection{Sensitivity to model parameter priors}
Calibration results may exhibit sensitivity to the choice of the model parameters' prior distributions \citep[cf.][]{Jackson2015,Reese2004}, especially for sparse observational records. This constitutes an important source of second-order, or deep uncertainty, an important factor in the design of risk management strategies \citep{keller2008dynamics}. To examine prior sensitivity, we calibrate the ice sheet model using two sets of prior distributions which are in the form of uniform or log-uniform distributions. One set of priors has a much wider range (large difference between upper and lower bounds) than the other. The much wider ranges represent physically possible parameter values that do not violate any fundamental physical laws, and the narrower ranges represent values that yield reasonable model behavior found in many years of unstructured tuning by the model developers \citep{pollard2012description}. We provide additional details in the Appendix. 

\begin{figure}[ht]
    \centering
    \includegraphics[width=1\textwidth]{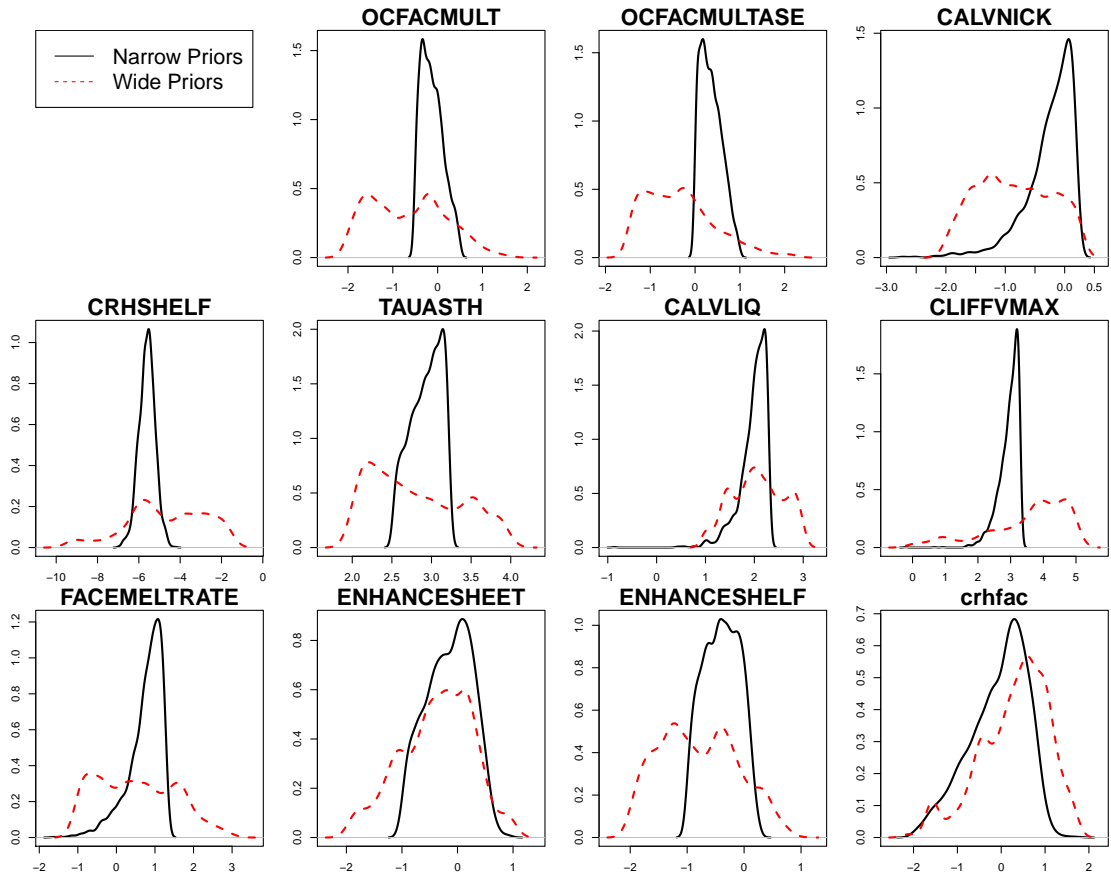}
    \caption{Posterior densities of model parameters using expert prior distributions (solid black lines) and wider expert prior distributions (dashed red lines). The dissimilarity of posterior distributions indicate that calibration results are highly sensitive to the choice of prior distributions.}
    \label{Fig:PriorSensPar}
\end{figure}

\begin{figure}[ht]
    \centering
    \includegraphics[width=1\textwidth]{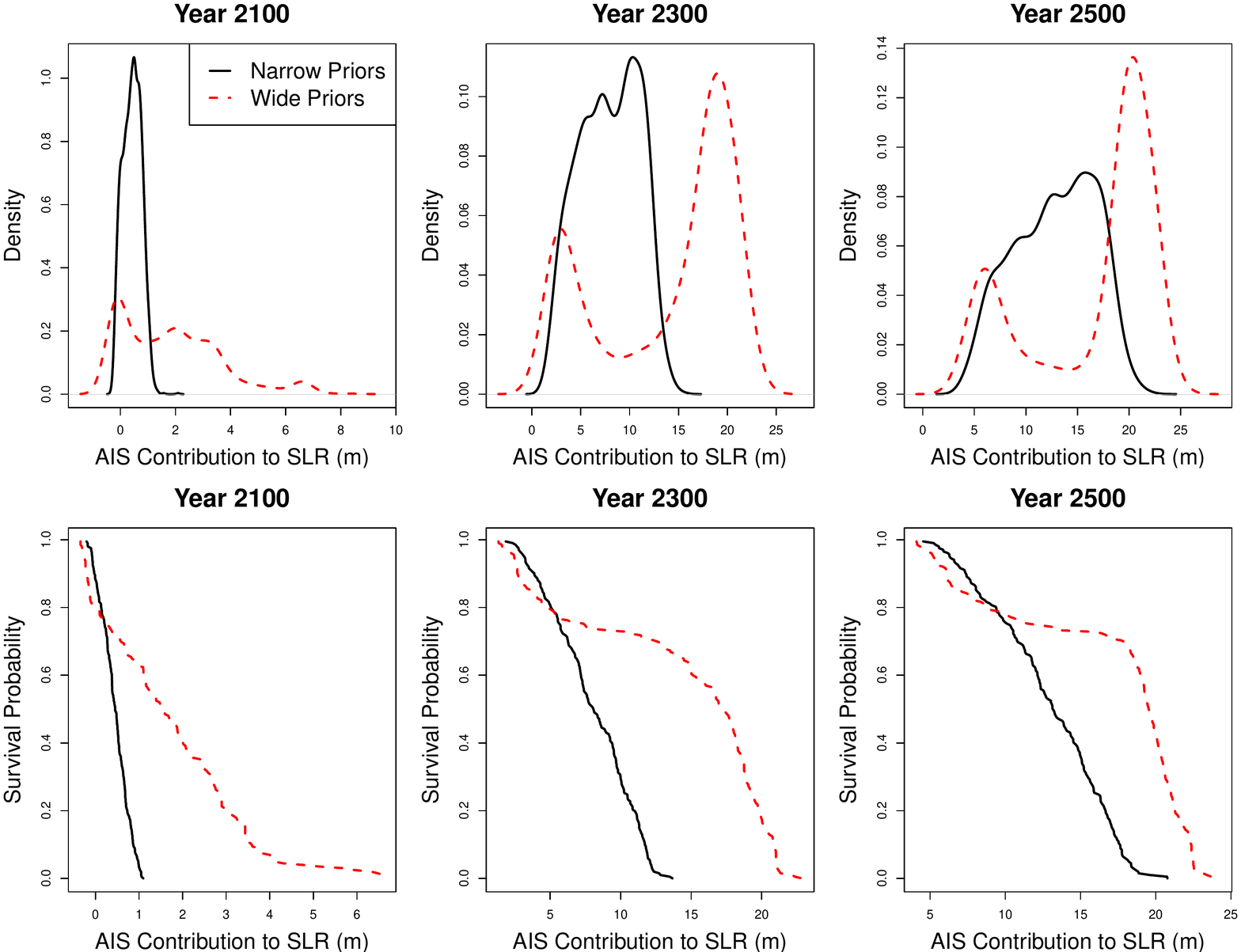}
    \caption{(Top Panel) Posterior densities of the projected Antarctic ice sheet's contribution to sea level change in 2100, 2300, and 2500 using expert prior distributions (solid black lines) and wider expert prior distributions (dashed red lines). (Bottom Panel) Empirical survival function of the projected Antarctic ice sheet's contribution to sea level change in 2100, 2300, and 2500 using expert prior distributions (solid black lines) and wider expert prior distributions (dashed red lines). For wide prior distributions, projections for future sea level rise is higher and more uncertain, and there exists bi-modality in the projections' posterior predictive distribution.}
    \label{Fig:PriorSensProj}
\end{figure}

The choice of prior distributions has a notable effect on parameter estimates (Figure \ref{Fig:PriorSensPar}) and sea level projections (Figure \ref{Fig:PriorSensProj} and Table \ref{Table:PriorSens}). Note that constraining the model parameters \textit{a priori} may underestimate sea level projections. However, overly wide prior distributions may permit physically unrealistic outcomes. Hence, it is important to carefully construct prior distributions based on domain area expertise, as we have in this manuscript. In particular, changing the prior on the parameter CLIFFVMAX -- wastage rate for unstable marine ice cliffs -- can have a strong impact on projections. For a prior range of 0 km/year to 12 km/year, the 95\% credible interval for the Antarctic ice sheet's contribution to sea level rise in 2300 is 1.2 m to 12.4 m. A wider prior range of 0 km/year- to 600 km/year results in considerably higher projection uncertainty denoted by a 95\% credible interval of 0.7 m to 21.0 m. 

\begin{table}[ht]
\caption{Antarctic ice sheet's projected contribution to sea level change in 2100-2500 after calibration using narrow and wide prior distributions.}
\centering
\begin{tabular}{cccccc}
  \hline
Prior & Year 2100 & Year 2200 & Year 2300 & Year 2400 & Year 2500 \\ 
  \hline
Narrow & 0.4 (-0.3, 1) & 3.8 (0.1, 6.7) & 7.9 (1.2, 12.4) & 10.6 (2.5, 15.5) & 12.8 (3.7, 18.5) \\ 
  Wide & 1.8 (-0.4, 5.5) & 10 (-0.2, 19.5) & 13.9 (0.7, 21) & 15.5 (1.8, 21.8) & 16.6 (3.1, 22.5) \\ 
   \hline
\end{tabular}
\label{Table:PriorSens}
\end{table}

 The wider range for CLIFFVMAX explores a fundamental uncertainty in MICI – the rate at which very tall ice cliffs will disintegrate back into the ice sheet interior. If grounding lines retreat into the interior of deep Antarctic basins, the exposed ice cliffs will be taller than any observed today, and the wastage velocities (CLIFFVMAX) could conceivably be much greater than the approximately 12 km per year observed today at the ice fronts of major Greenland glaciers (which might not even be approximate analogs for MICI, being driven instead mainly by buoyant calving; \citet{murray2015dynamics}). The bimodal character of the posterior densities in the top panels of Figure \ref{Fig:PriorSensProj} for 2300 and 2500 are due to the very large CLIFFVMAX range. The upper peak centered on around 20 m is produced by CLIFFVMAX values of approximately 100 km per year and above, which produce collapse of almost all marine ice in both East and West Antarctica. The lower peak centered on around 5 m occurs for many lower CLIFFVMAX values, for which the more vulnerable West Antarctic ice sheet collapses, but marine basins in East Antarctica do not retreat.

\section{Discussion}\label{Sec:Discussion}
\subsection{Summary}
We present a novel particle-based approach to calibrate the $80$ km resolution PSU3D-ICE model. We show that our approach provides good approximations and drastically reduces overall calibration wall times by heavily parallelizing the sequential Monte Carlo algorithm, and carefully tuning the algorithm to drastically reduce the number of sequential model evaluations. Our algorithm is applicable to a broad class of models that have a moderate run time (given our computing resources, between a few seconds and several minutes) and a moderate number of model parameters (in our case between 5 and 20). 

We use this new method to assess the impacts of neglecting parametric uncertainties on sea level projections. Emulation-calibration methods using fewer parameters yield lower and more overconfident projections of sea level rise than using more parameters through the particle-based calibration approach. This method includes the recent study of \citet{Edwards2019}, who found that the important mechanism of marine ice cliff instability (MICI) is not necessary to capture past variations. In this case, future sea level projections are considerably lower. In contrast, our new approach that accounts for more parametric uncertainties suggests that MICI may still be important and future sea level projections may be much higher, especially considering potential Pliocene windows. Using emulation-calibration in a high-dimensional parameter space induces considerable emulator-model discrepancy and can result in large projection uncertainties. Our method utilizes the actual ice sheet model; thereby preserving the highly non-linear ice dynamics as well as the complex interactions between model parameters. This has clear policy-relevant implications because projections from ice sheet models inform economic and engineering assessments \citep[cf.][]{sriver2018characterizing, diaz2016potential, johnson2013estimating}. 

Our approach enables calibration experiments that were computationally prohibitive using current calibration methods. First, assuming different ranges of Pliocene era sea level constraints (low vs. high) results in markedly different characterizations of parametric uncertainty and projections of sea level rise over the next five centuries. These results suggest that improved geological data from the Pliocene can help better quantify the model parameters central to marine ice cliff instability (MICI) and improve sea level projections. 
Second, calibration results are highly sensitive to the choice of prior distributions. Over-constraining the prior distributions (in particular by not allowing very fast cliff disintegration rates), we can mischaracterize parametric uncertainty and drastically underestimate future sea level changes.

\subsection{Caveats}
Our conclusions are subject to the usual caveats that also point to promising and policy-relevant research directions. Key methodological caveats include that our calibration approach may not scale well to computer models with long model run times ($>15$ minutes), high-dimensional input spaces ($>20$ parameters), or a combination of both. Our approach not be suitable for computer models that use multiple processors for a single model run. 
Selecting an appropriate number of particles remains an open question. Past theoretical work \citep{crisan2000convergence} state that using more particles yields better approximations of the target distributions. Here, we set the total particle count with respect to the available resources. 
  
A number of caveats apply to our scientific findings. Using the PSU3D-ICE model at a coarser resolution than previous studies \citep{deconto2016contribution,chang2016calibrating,chang2016improving, Pollard2016Large} is admittedly a compromise between physical fidelity and run-time feasibility. At coarser resolutions, complex ice processes may not properly coalesce due to the spatial constraints. 
As discussed above in Section \ref{Sec:PSU3DICE}, this model is well suited to coarse-resolution studies because important sub-grid processes are parameterized without explicit grid dependence, and previous sensitivity tests have shown that results are reasonably independent of resolution. Nevertheless, replicating this calibration study at sharper spatial resolutions (40 km to 10 km) is a natural and worthwhile extension of this study. Promising avenues for future work would include incorporating parallel MCMC approaches such as Multiple-Try Metropolis \citep{liu2000multiple} or ``emcee'' samplers \citep{goodman2010ensemble} to reduce computer model runs in the mutation stage.
Finally, the likelihood functions for the paleoclimate records may heavily influence calibration results. We have shown how the choice of expert priors influence calibration, but the influence of likelihood functions remains unexamined. 

\subsection*{Acknowledgements}
We would like to thank Don Richards, Daniel Gilford, Bob Kopp, Kelsey Ruckert, Vivek Srikrishnans, Robert Ceres, Kristina Rolph, Mahkameh Zarekarizi, and Casey Hegelson for useful discussions. This study was partially supported by the Department of Energy sponsored Program on Coupled Human Earth Systems (PCHES) under DOE Cooperative Agreement No.DE-SC0016162 and by the National Science Foundation through the Network for Sustainable Climate Risk Management (SCRiM) under NSF cooperative agreement GEO-1240507. This study was also co-supported by the Penn State Center for CLimate Risk Management. We would like to acknowledge high-performance computing support from Cheyenne (doi:10.5065/D6RX99HX) provided by NCAR's Computational and Information Systems Laboratory, sponsored by the National Science Foundation. Any opinions, findings, and conclusions or recommendations expressed in this material are those of the authors and do not necessarily reflect the views of the Department of Energy, the National Science Foundation, or other funding entities. Any errors and opinions are, of course, those of the authors. We are not aware of any real or perceived conflicts of interest for any authors.

\subsection*{Author contributions}

All authors co-designed the overall study. BSL and MH formulated the statistical method. BSL wrote the computer code for calibration, designed the Pliocene window analysis, and wrote the first draft of the manuscript. RF integrated the calibration method into the Cheyenne high perfomance computing system. MH edited the text. KK designed the comparative methods analysis and edited the text. DP provided code and data for the $80$ km resolution PSU3D-ICE model, designed the prior sensitivity study, and edited the text. 

\subsection*{Code and data availability}
All code, data, and output will be available under the GNU general public open-source license through the corresponding author, available via Github at \href{https://github.com/scrim-network/LeeEtal-PSUICE-3D}{https://github.com/scrim-network/LeeEtal-PSUICE-3D} upon publication. The results, data, software tools, and other resources related to this work are provided as-is without warranty of any kind, expressed or implied. In no event shall the authors or copyright holders be liable for any claim, damages or other liability in connection with the use of these resources.

\newpage
\bibliography{references}

\end{document}


\title{Supplement:\\
A Fast Particle-Based Approach for Calibrating a 3-D Model of the Antarctic Ice Sheet}
\author{Ben Seiyon Lee, Murali Haran, Robert Fuller, David Pollard, and Klaus Keller}
\maketitle

\section{Parameter Descriptions}
We calibrate 11 model parameters of the PSU3D-ICE model. The parameter descriptions are as follows:
\begin{enumerate}
    \item \textbf{OCFACMULT:} A dimensionless coefficient multiplying the rate of sub-oceanic melting or freezing calculated at the base of floating ice shelves \citep{Pollard2016Large, pollard2012description}. It corresponds to parameter $\kappa$ in equation 17 of \citet{pollard2012description}. The calculation of sub-ice-shelf melt rate primarily depends on the temperature of nearby oceanic water at 400 m beneath sea level \citep{pollard2012description}.
    
    \item \textbf{OCFACMULTASE:} A dimensionless coefficient that modifies the sub-oceanic ice shelf melting or freezing rate in the Amundsen Sea Embayment of the West Antarctic Ice Sheet \citep{chang2016calibrating}. Oceanic melting may occur at a different rate here due to stronger regional circulation \citep{jacobs_stronger_2011}.
    
    \item \textbf{CRHSHELF:} A dimensionless multiplier applied uniformly to basal sliding coefficients for continental shelf areas (modern ocean areas). It multiplies the basal sliding coefficients $C'$ in equation 10 of \citet{pollard2012description}, which have units of m $\text{year}^{-1}$ $\text{Pa}^{-2}$. 

    \item \textbf{CRHFAC: } A dimensionless multiplier applied uniformly to basal sliding coefficients for areas with modern grounded ice and was calculated previously using a simple inverse method\citep{pollard_simple_2012}. It multiplies the basal sliding coefficients $C'$ in equation 10 of \citet{pollard2012description}, which have units of m $\text{year}^{-1}$ $\text{Pa}^{-2}$. 

    \item \textbf{ENHANCESHEET: } A dimensionless coefficient multiplying the rheologic coefficient in the calculation of the viscous vertical-shearing deformation of ice. This calculation uses the shallow ice approximation (SIA), usually the dominant mode of flow for grounded ice. It corresponds to $E$ in equation 16 of \citet{pollard2012description}.

    \item \textbf{ENHANCESHELF: } A dimensionless coefficient multiplying the rheologic coefficient in the calculation of the viscous horizontal-stretching deformation of ice. This calculation uses the shallow shelf approximation (SSA), usually the dominant mode of flow for floating ice. It corresponds to $E$ in equation 16 of \citet{pollard2012description}.
    
    \item \textbf{FACEMELTRATE: } A dimensionless coefficient multiplying the melt rate of vertical ice cliffs in contact with warm ocean water at the edges of ice shelve \citep{pollard2012description}.

    \item \textbf{TAUASTH: } The e-folding time, for local asthenospheric relaxation towards isostatic equilibrium, in the calculation of bedrock response to varying ice loading and unloading.  Units are in years, and it corresponds to $\tau$ in equation 33 of \citet{pollard2012description}.
    
    \item \textbf{CLIFFVMAX: } The maximum erosional retreat rate for unstable marine ice cliffs exceeding approximately 100 meters in sub-aerial height \citep{pollard2015potential}. This is the horizontal material velocity of cliff wastage into the upstream solid ice, in the parameterization of  marine ice cliff instability (MICI). Units are in meters per year.
    
    \item \textbf{CALVLIQ:} Scaling depth for the deepening of surface crevasses by hydrofracturing due to surface melt and rainfall. Its units are meters of crevasse depth and is the crevasse deepening produced by a surface melt plus rainfall rate of 1 meter per year. It corresponds to the constant 100 in equation B.6 of \citet{pollard2012description}.
    
    \item \textbf{CALVNICK : } A dimensionless coefficient multiplying the combined total depth of crevasses in the calving parameterization. This depth is compared to the actual ice-shelf thickness in the model’s calving parameterization \citep{pollard2015potential, nick_physically_2010}. It multiplies the parameter $r$ in equation B.7 of \citet{pollard2012description}.

\end{enumerate}

\begin{figure}[p]\label{SuppFig:Map}
    \centering
    \includegraphics[width=1\textwidth]{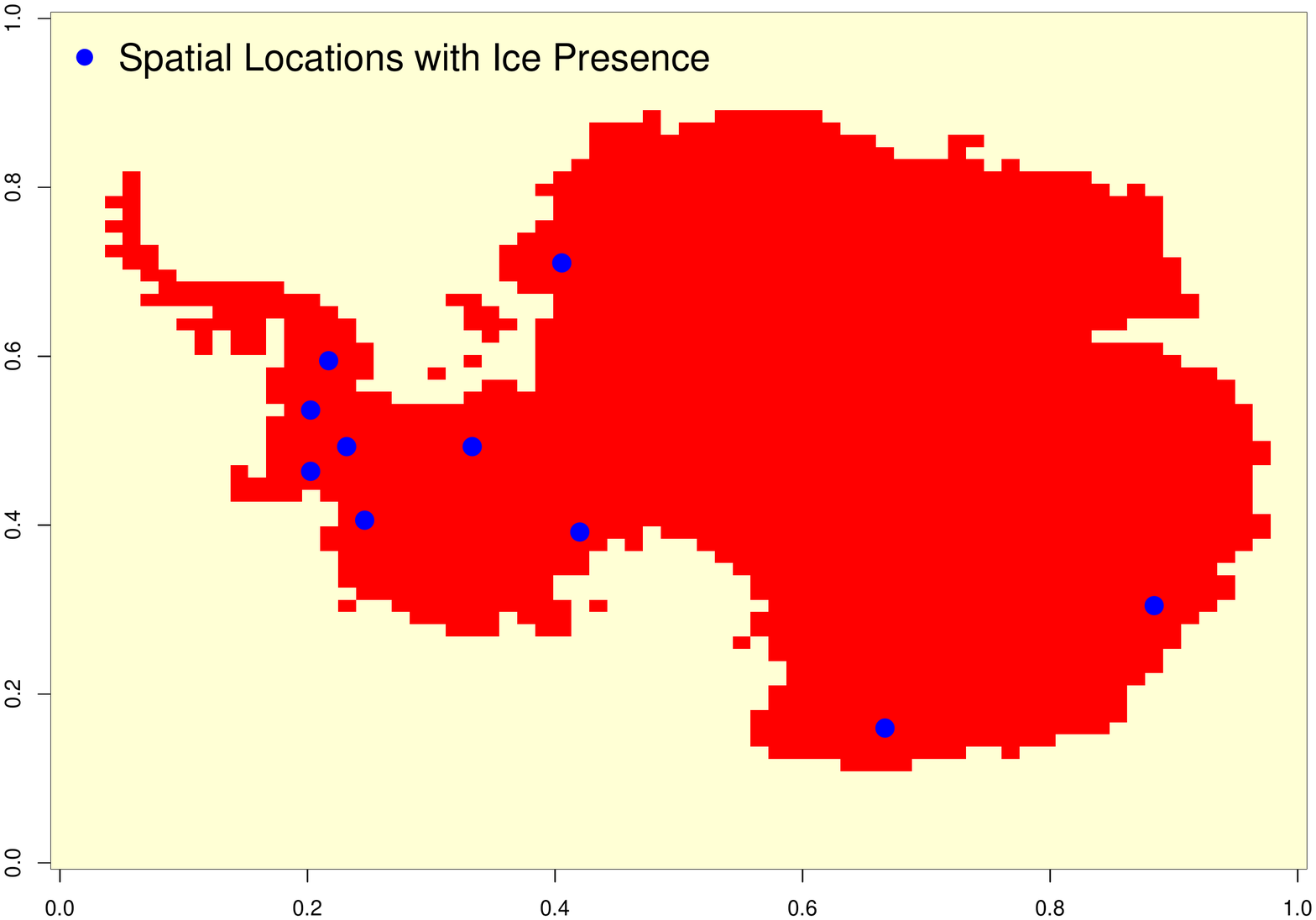}
    \caption{Modern observations of ice presence obtained via the Bedmap2 project. The blue dots indicate locations where there is confirmed ice presence.}
\end{figure}

\section{Simulated Example}
\begin{figure}[h]
    \centering
    \includegraphics[width=1\textwidth]{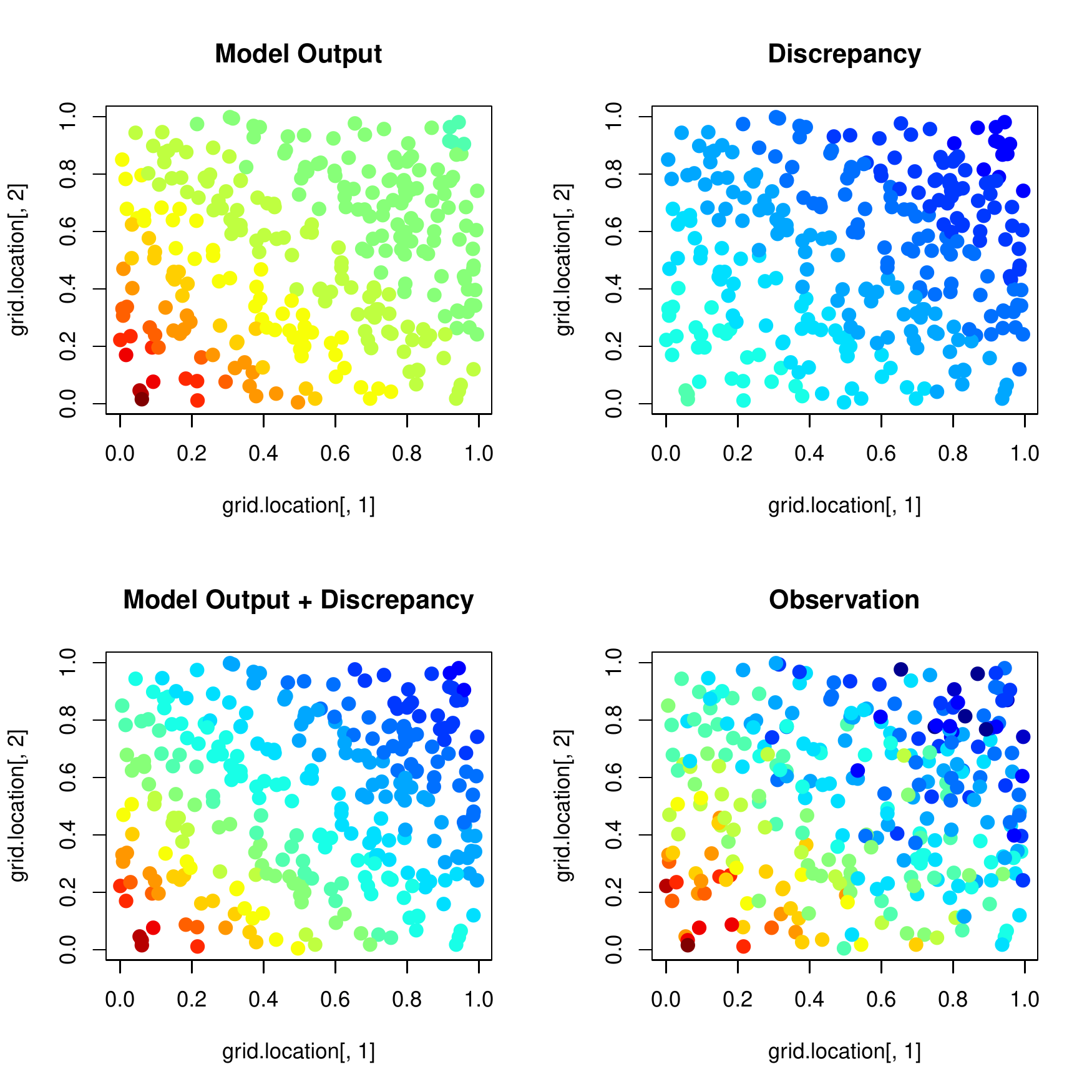}
    \caption{(Top left) Map of the model output from the toy example. (Top right) Map of the systematic and also spatially correlated data-model discrepancy. (Bottom left) Map of the sum of the model output and discrepancy.  (Bottom right) Map of the observations, which is the sum of the model output, discrepancy, and iid observational error. }
    \label{SuppFig:ToyData}
\end{figure}

\begin{figure}[h]
    \centering
    \includegraphics[width=1\textwidth]{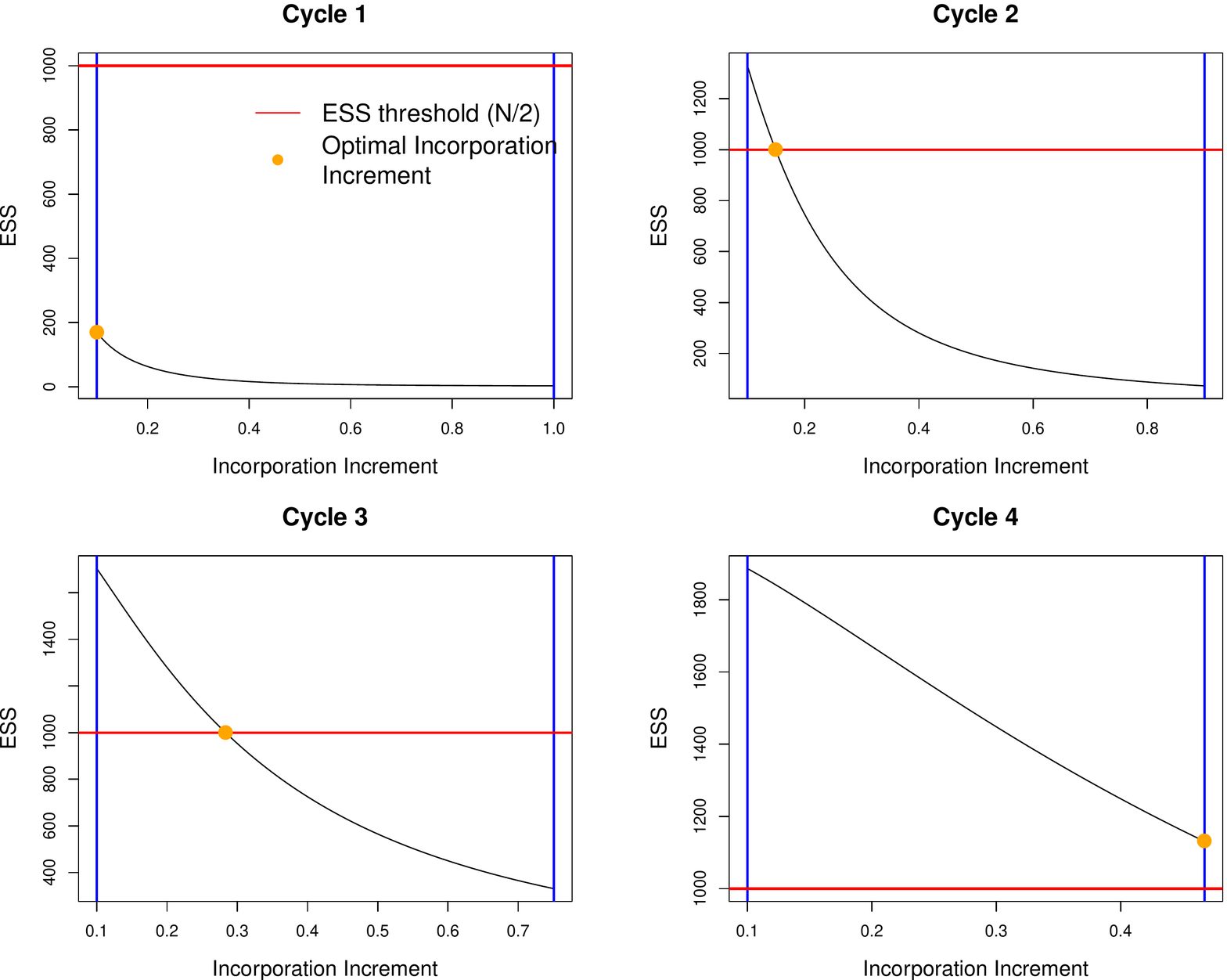}
    \caption{Incorporation increment $\gamma_t$ selection for the simulated example. Each panel corresponds to a cycle (4 total). The x-axis denotes possible values for the incorporation increment $\gamma_t$ and the y-axis denotes the corresponding effective sample size (ESS). The red line represents the ESS threshold set at $N/2$. The orange point denotes the optimal incorporation increment and the corresponding ESS at each cycle.}
    \label{SuppFig:ToyGamma}
\end{figure}

\begin{figure}[h] 
    \centering
    \includegraphics[width=1\textwidth]{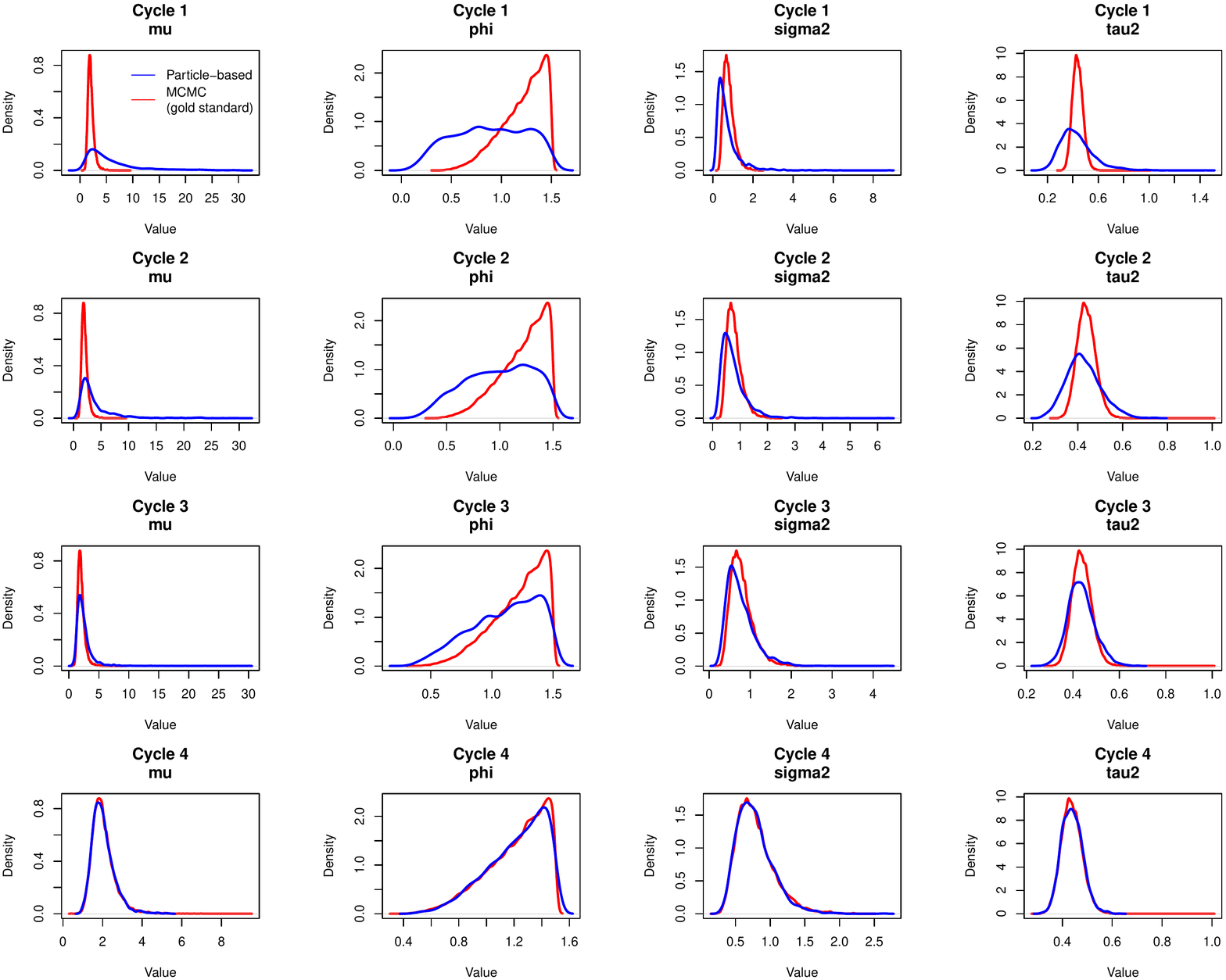}
    \caption{Posterior densities for the simulated example after each cycle. Each row corresponds to a cycle, and each column corresponds to a model parameter. The blue lines represent the density of the posterior samples from the particle-based approach, and the red lines denote the density of the posterior samples obtained form MCMC (gold standard). Note that the particle-based approach provides a good approximation to the MCMC-based approach. However, the particle-based approach requires just 40 model evaulations as opposed to 100k for the MCMC-based approach. }
    \label{Fig:ToyDensity}
\end{figure}

We provide additional details pertaining to the simulation study using $N=2000$ particles from Section 5. Maps of the model outputs and observations are provided in Figure \ref{SuppFig:ToyData}. The simulated calibration experiment went through four sampling-importance-resampling cycles with corresponding incorporation increments $\gamma=\{0.100, 0.15, 0.27, 0.47\}$. Our adaptive likelihood incorporation schedule chose four sampling-importance-resampling cycles. In the first cycle, our algorithm chose a incorporation increment $\gamma_{1}=0.1$, which yields an effective sample size (ESS) of $169.5$. In the second cycle, the algorithm chose an incorporation increment $\gamma_{2}=0.15$ with a corresponding ESS of $1000$. For the third cycle, the selected incorporation increment is $\gamma_{3}=0.27$ with a corresponding ESS of $1000$. In the fourth and final cycle, we use an incorporation increment of $\gamma_{4}=0.47$ with a corresponding ESS of $1143$. Figure \ref{SuppFig:ToyGamma} shows the chosen incorporation increments and corresponding ESS for each cycle. Figure \ref{Fig:ToyDensity} displays posterior parameter densities after each cycle. 

In the mutation stage, we chose the baseline number of Metropolis-Hastings updates to be five updates. Our algorithm determined that the empirical distribution of our stopping metric, model parameter $\theta$, stabilizes after 10 total iterations. The stopping criterion is met once the Batthacharyya distance of the empirical samples at the 10-th mutation update and the 5-th update is less than a pre-determined threshold. 

\section{Emulation-Calibration Details}
We provide additional details regarding the comparative analysis performed in Section 5.3.  Available paleoclimate and observational data include the Antarctic ice sheet's contribution to sea level change in the (1) Pliocene era; (2) Last Interglacial Age; (3) Last Glacial Maximum; (4) the total volume of the Antarctic ice sheet in the modern era; and (5) total grounded area of the Antarctic ice sheet in the modern era. For the comparison study, we omit the binary observations (ice vs. no ice) obtained at the 10 strategic locations (Manuscript Section 6.1). We use the same prior distributions for our model parameters as provided in Section 6.1 of the manuscript. 

For three-parameter emulation-calibration example, we select OCFACMULT, CALVLIQ, and CLIFFVMAX as the calibration parameters and fix the remaining eight parameters. To train the Gaussian process emulator, we use PSU3D-ICE output obtained at 512 different input parameter settings. We generate the input parameter settings using a full factorial design, which includes eight discrete levels for each model parameter. The eight levels span the uniform prior distribution ranges as provided in Section 6.1 of the manuscript. We fit a separate Gaussian process for each modern and paleo-climate observational record (5 total); in addition, we fit a Gaussian process for the Antarctic ice sheet contribution to sea level change in 2100, 2200, 2300, 2400, and 2500. Each Gaussian process has the form $Y(\theta)\sim \mathcal{GP}(\mu(\theta;\beta_{0};\beta),C(\theta,\theta';\sigma^{2},\phi))$, where the mean function $\mu(\theta;\beta_{0};\beta)=\beta_{0}+\beta\theta$ includes an intercept and a linear trend. We use a squared exponential covariance function, $C(\theta,\theta';\sigma^{2},\phi) = \sigma^{2}\prod_{i=1}^{p}\exp\{-\frac{(\theta_{i}-\theta_{i}')^2}{\phi_{i}}\}$, where $\theta\in\mathbb{R}^{p}$ $\phi=(\phi_{1},...,\phi_{p})$. We estimate the Gaussian process parameters, $(\beta, \sigma^{2}, \phi)$, through maximum likelihood estimation. we fit the Gaussian process emulator using the \textbf{mlegp} R package \citep{mlegp2008}. The 3-parameter Gaussian process emulator has a low out-of-sample cross validated root mean squared prediction error as shown in Table \ref{Table:11GPE}.

In the 11-parameter emulation-calibration study, we implement a two-part emulation-calibration method using all model parameters. We run the PSU3D-ICE model at 512 input parameter settings chosen through a Latin Hypercube Design (LHC). The LHC samples span the ranges of the prior distributions provided in Section 6.1 of the manuscript. Similar to the three-parameter case, we fit a Gaussian process emulator via maximum likelihood estimation. The 11-parameter Gaussian process emulator has a high out-of-sample cross validated root mean squared prediction error, as shown in Table \ref{Table:11GPE}. This can be attributed to the low-fidelity emulator trained using a small number of design points (512) to explore an 11-dimensional parameter space.

\begin{table}[ht] \label{Table:11GPE}
\centering
\begin{tabular}{ccc}
  \hline
 & 3 Parameter Emulator & 11 Parameter Emulator \\ 
 & RMSE & RMSE \\ 
  \hline
Pliocene & 0.20 & 1.08 \\ 
  Last Interglacial & 0.15 & 0.87 \\ 
  Last Glacial Maximum & 0.02 & 6.18 \\ 
  Modern SLE & 0.25 & 7.02 \\ 
  Modern Volume & 0.18 & 3.73 \\ 
  Year 2100 & 0.27 & 5.71 \\ 
  Year 2200 & 0.37 & 6.40 \\ 
  Year 2300 & 0.23 & 1.92 \\ 
  Year 2400 & 0.26 & 0.87 \\ 
  Year 2500 & 0.23 & 0.82 \\ 
   \hline
\end{tabular}
\caption{Out-of-sample cross validated root mean squared prediction error (RMSE) for a Gaussian process emulator with 3 parameters and 11 parameters. The three-parameter emulator exhibits low RMSE across all observations and projections. The 11-parameter emulator has a high RMSE, which is indicative of a low-fidelity, or inaccurate, surrogate model.}
\end{table}

\section{Prior Sensitivity Analysis}
We conduct a prior sensitivity analysis using two sets of prior distributions provided by domain experts. The first set of prior distributions are from the main calibration experiment in Section 6.1 of the manuscript. The second set of prior distributions includes extended ranges for the model parameters. Note that we change the prior distribution for model parameters -- CALVNICK, TAUASTH, CALVLIQ, CLIFFVMAX, FACEMELTRATE -- from a uniform distribution to a log-uniform distribution. The second set of prior distributions are: 
\begin{multicols}{2}
\begin{itemize}
    \item $\log_{10}(\theta_{OCFACMULT})\sim \mathcal{U}(-2, 2)$
    \item $\log_{10}(\theta_{OCFACMULTASE})\sim \mathcal{U}(-1.5, 2.5)$
    \item $\log_{10}(\theta_{CALVNICK})\sim \mathcal{U}(-2, 2)$
    \item $\log_{10}(\theta_{CRHSHELF})\sim \mathcal{U}(-9.5, -1.5)$
    \item $\log_{10}(\theta_{CALVLIQ})\sim \mathcal{U}(1, 3)$
    \item $\log_{10}(\theta_{FACEMELTRATE})\sim \mathcal{U}(-1, 3)$
    \item $\log_{10}(\theta_{ENHANCESHEET})\sim \mathcal{U}(-2, 2)$
    \item $\log_{10}(\theta_{CRHFAC})\sim \mathcal{U}(-3, 3)$
    \item $\log_{3}(\theta_{TAUASTH})\sim \mathcal{U}(2, 4)$
    \item $\log_{6}(\theta_{CLIFFVMAX})\sim \mathcal{U}(0, 5)$
    \item $\log_{0.3}(\theta_{ENHANCESHELF})\sim \mathcal{U}(-2, 2)$
\end{itemize}
\end{multicols}

\begin{figure}[p]
    \centering
    \includegraphics[width=1\textwidth]{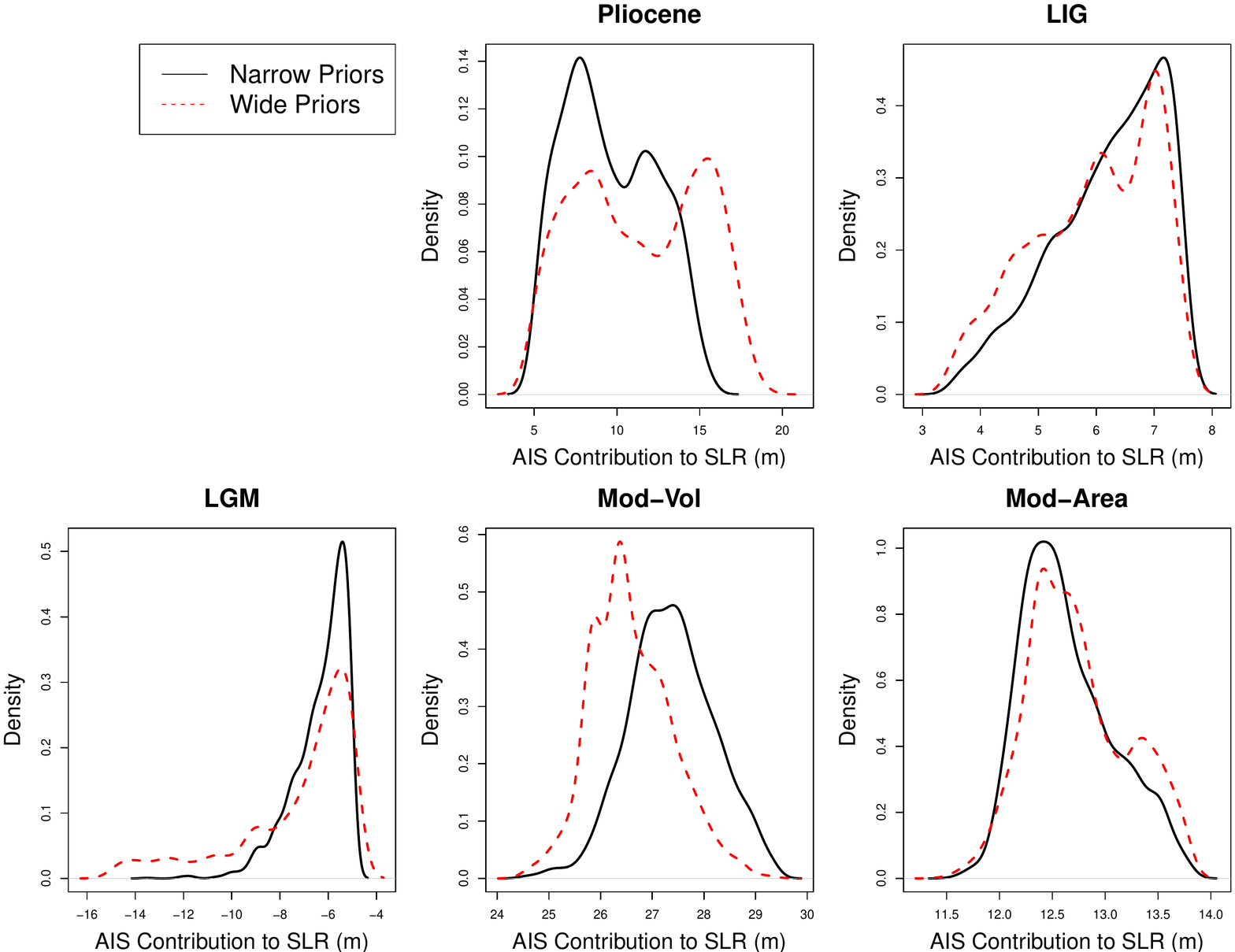}
    \caption{Posterior densities of observational records using expert prior distributions (solid black lines) and wider expert prior distributions (dashed red lines). Wider expert priors result in a bi-modal distribution for the AIS contribution to sea level rise in the Pliocene and lower modern volume, both in point estimate and 95\% credible intervals. }
    \label{SuppFig:WidePriorObs}
\end{figure}

\begin{figure}[p]
    \centering
    \includegraphics[width=1\textwidth]{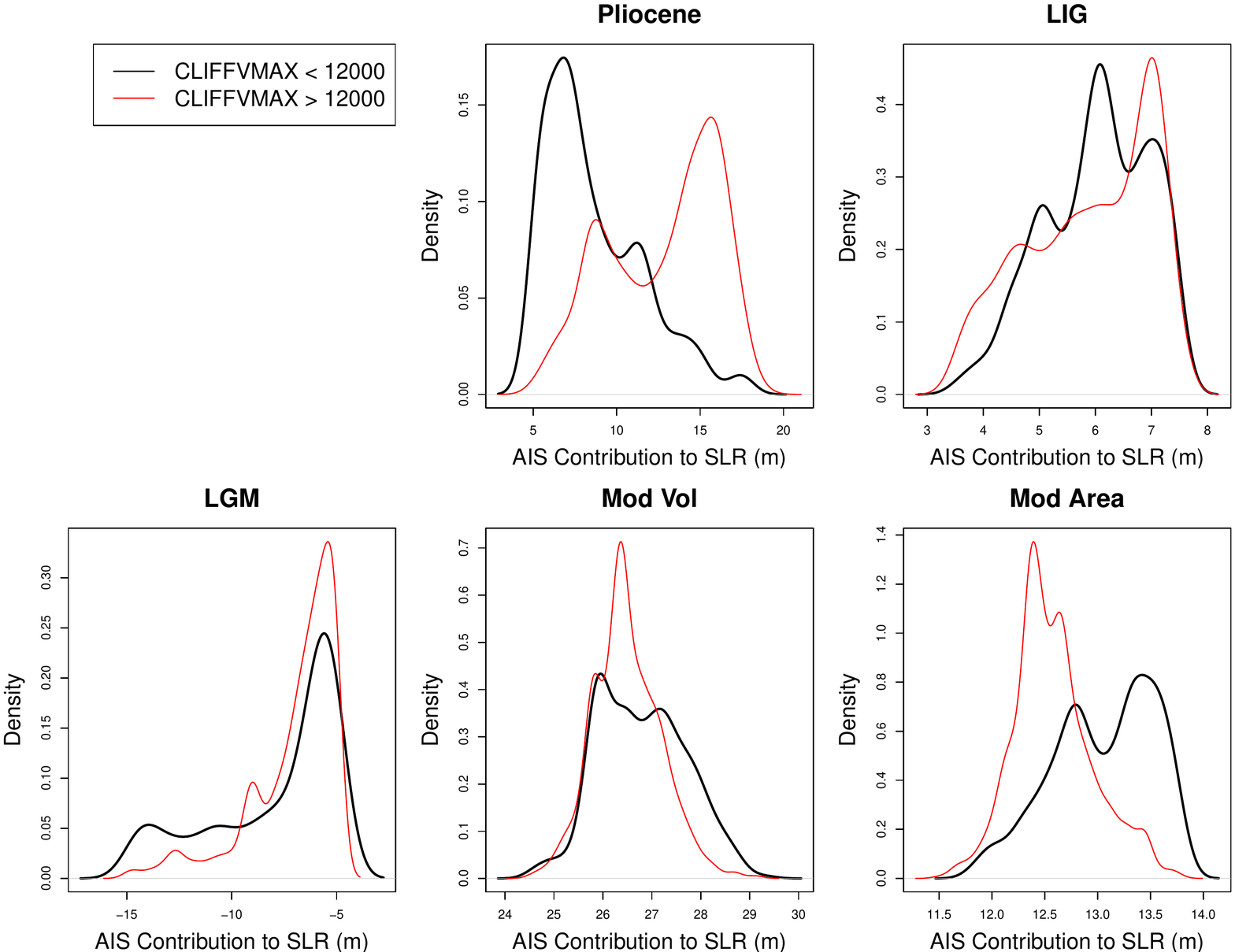}
    \caption{Posterior densities of observational records using the wider expert prior distributions. The posterior densities are split for values of CLIFFVMAX less than 12 km per year (black lines) and greater than 12 km per year (red lines). Higher values of CLIFFVMAX results in higher values (point estimates and 95\% credible intervals) of the the Antarctic ice sheet's contribution to sea level rise in the Pliocene and lower modern volume.}
    \label{SuppFig:CLIFFVMAXObs}
\end{figure}

\begin{figure}[p]
    \centering
    \includegraphics[width=1\textwidth]{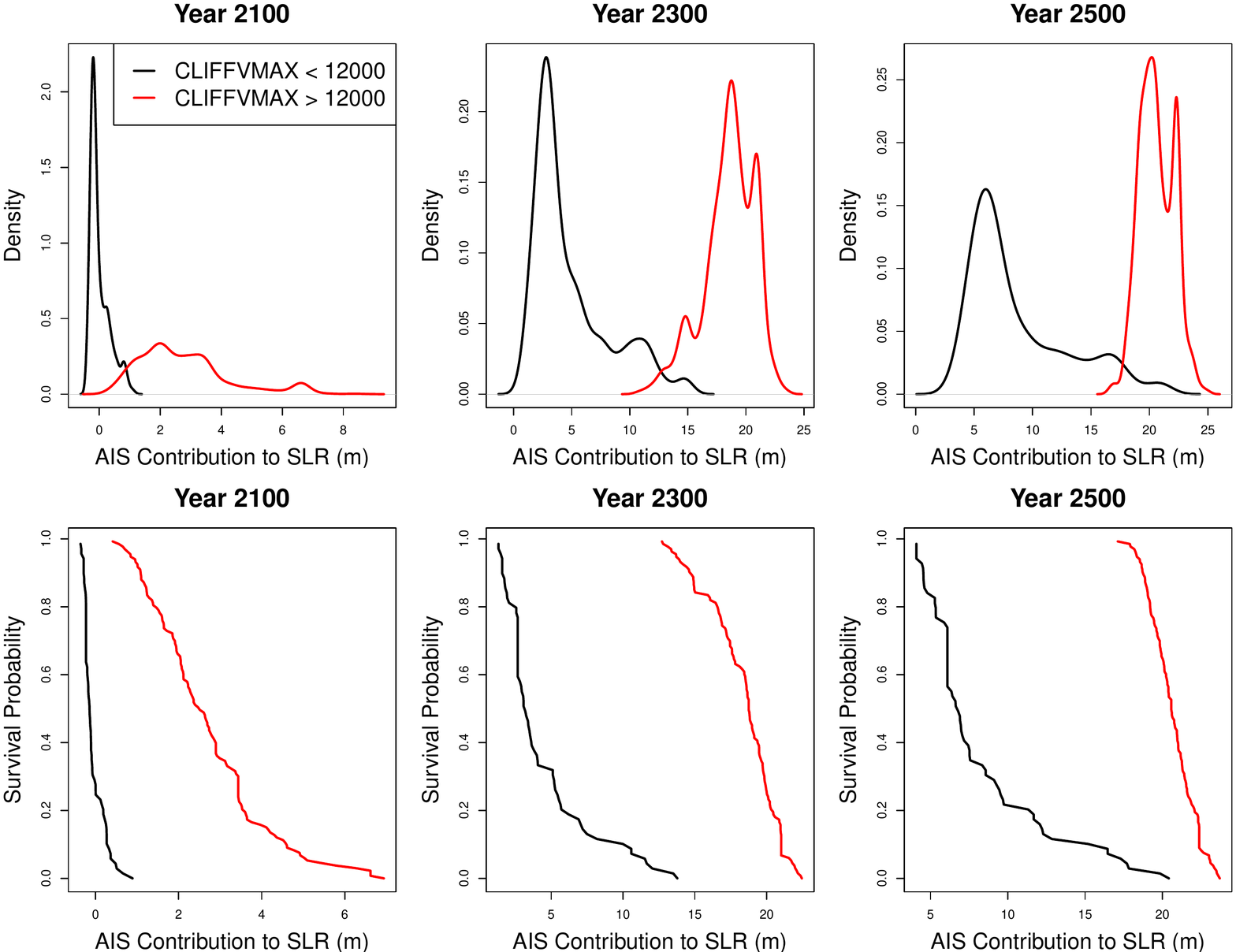}
    \caption{(Top Panel) Posterior densities of the projected Antarctic ice sheet's contribution to sea level change in 2100, 2300, and 2500 using the wider expert prior distributions. The posterior densities are split for values of CLIFFVMAX less than 12 km per year (black lines) and greater than 12 km per year (red lines). (Bottom Panel) Empirical survival function of the projected Antarctic ice sheet's contribution to sea level change in 2100, 2300, and 2500 for higher CLIFFVMAX values (solid black lines) and lower CLIFFVMAX values (red lines). Larger values of CLIFFVMAX results in considerably higher projections of future sea level rise, both in point estimates and 95\% credible intervals. }
    \label{SuppFig:CLIFFVMAXSurvival}
\end{figure}

\section{Comparison with Standard Particle-based Calibration}
One major contribution of this study is reducing the number of sequential likelihood evaluations. Each likelihood evaluation requires a computer model runs, which are the dominant costs of our approach. We introduce an adaptive likelihood incorporation schedule which is automated. In a standard implementation of the particle-based method, we must set the total number of sampling-importance-resampling cycles and the total number of mutation runs per cycle. Here, we compare results from a standard implementation to those using our fast adaptive method. In the standard implementation, we set the total number of cycles to be 6 and the total number of Metropolis-Hastings updates (for the mutation stage) to be 45. These chosen values are based on the available computing resources, namely a 12-hour walltime limit for each mutation cycle. In this comparison study, we use the five modern and paleoclimate records as observations; spatial constraints are omitted. 

\begin{figure}[p]
    \centering
    \includegraphics[width=1\textwidth]{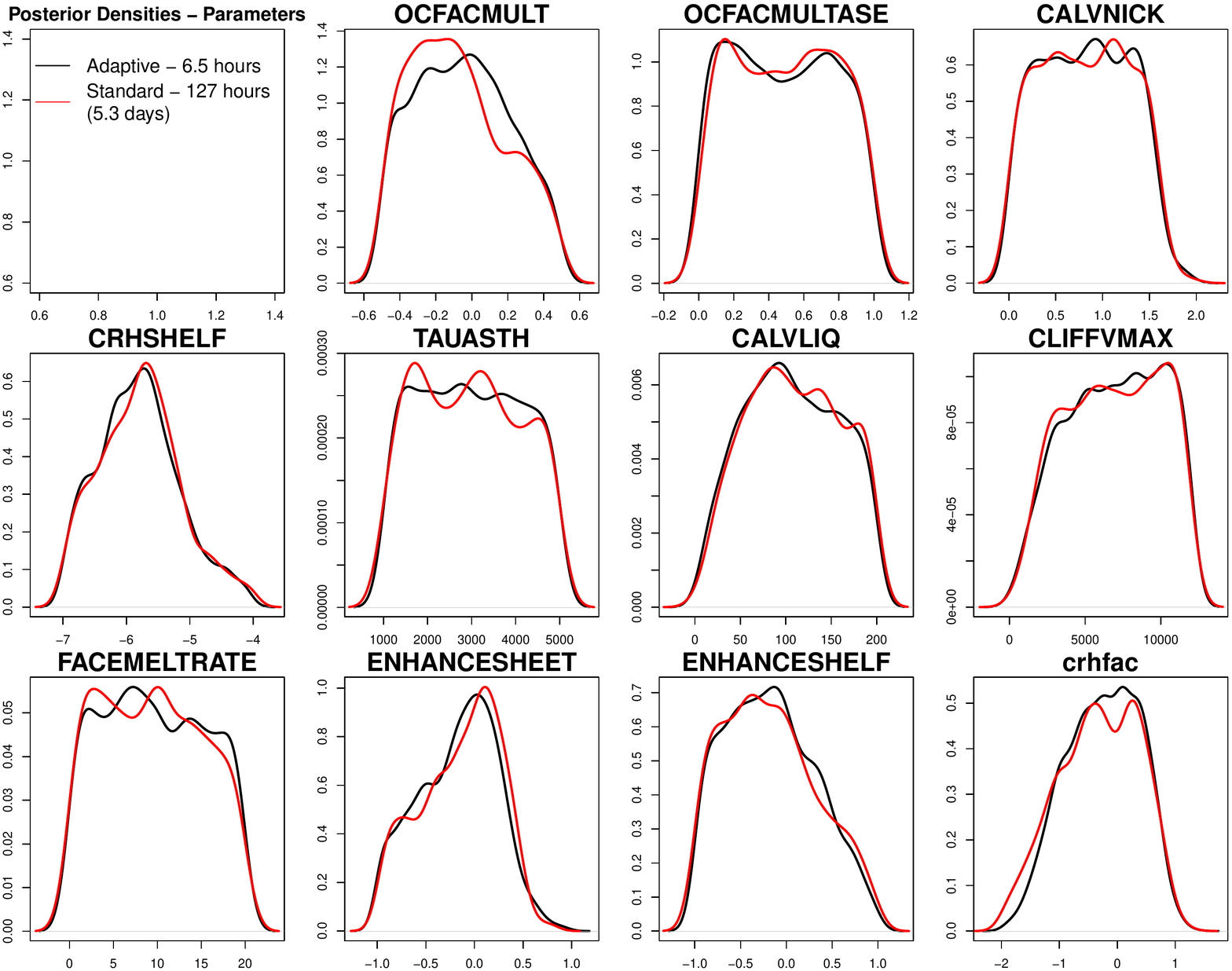}
    \caption{Posterior densities of model parameters using adaptive particle-based approach (solid black lines) and the standard particle-based approach (dashed red lines). The adaptive particle-based approach goes through 4 cycles and runs 14 updates in the mutation stage with a total calibration wall time of 6.5 hours. The standard particle-based approach goes through 10 cycles and runs 45 updates in the mutation stage with a total calibration wall time of 127 hours (5.3 days). Posterior densities for both methods are comparable.}
\label{SuppFig:StandardAdaptivePar}
\end{figure}

\begin{figure}[p]
    \centering
    \includegraphics[width=1\textwidth]{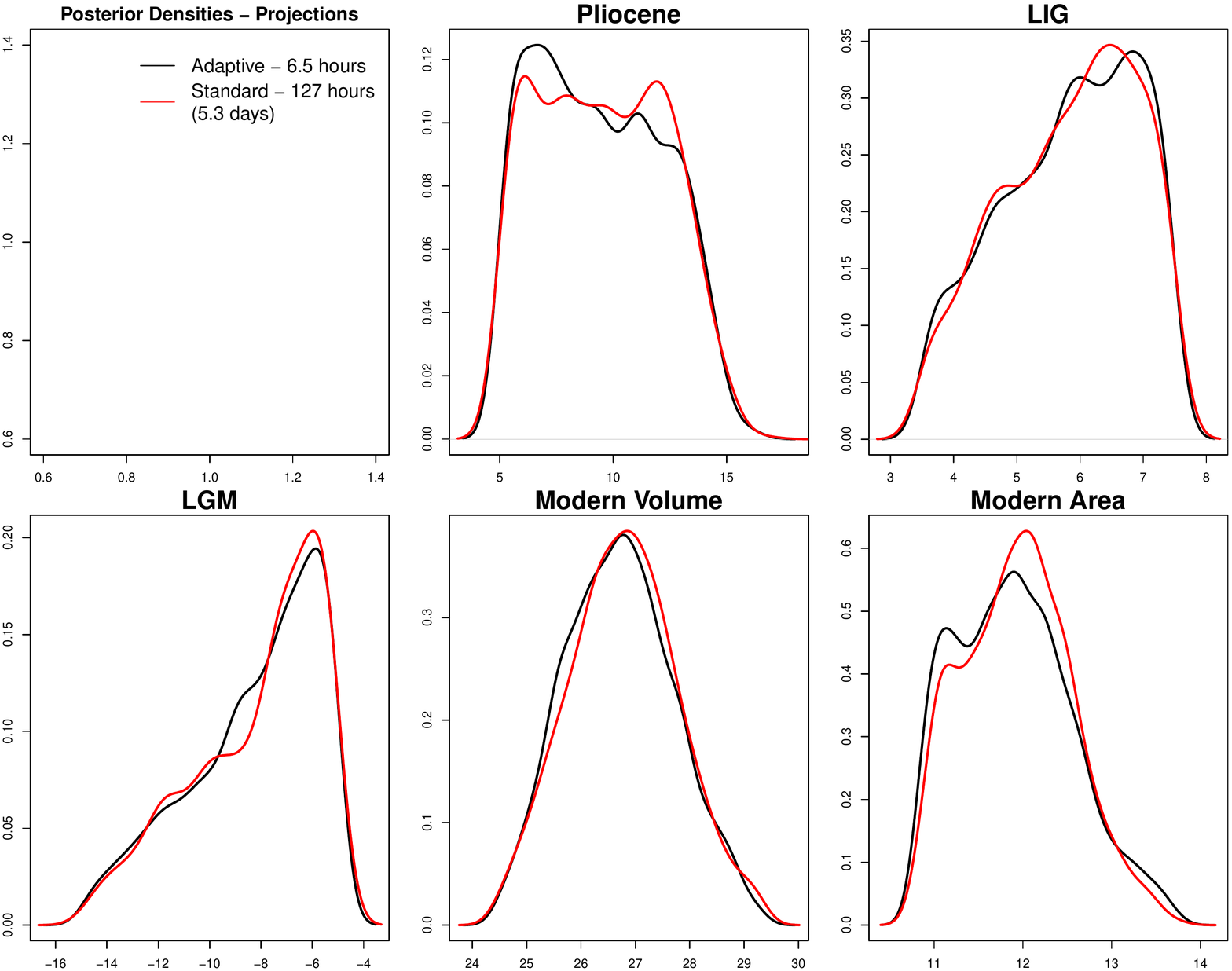}
    \caption{Posterior densities of observational records using adaptive particle-based approach (solid black lines) and the standard approach (dashed red lines). The adaptive particle-based approach goes through 4 cycles and runs 14 updates in the mutation stage with a total calibration wall time of 6.5 hours. The standard particle-based approach goes through 10 cycles and runs 45 updates in the mutation stage with a total calibration wall time of 127 hours (5.3 days). Posterior densities for both methods are comparable.}
\label{SuppFig:StandardAdaptiveObs}
\end{figure}

\begin{figure}[p]
    \centering
    \includegraphics[width=1\textwidth]{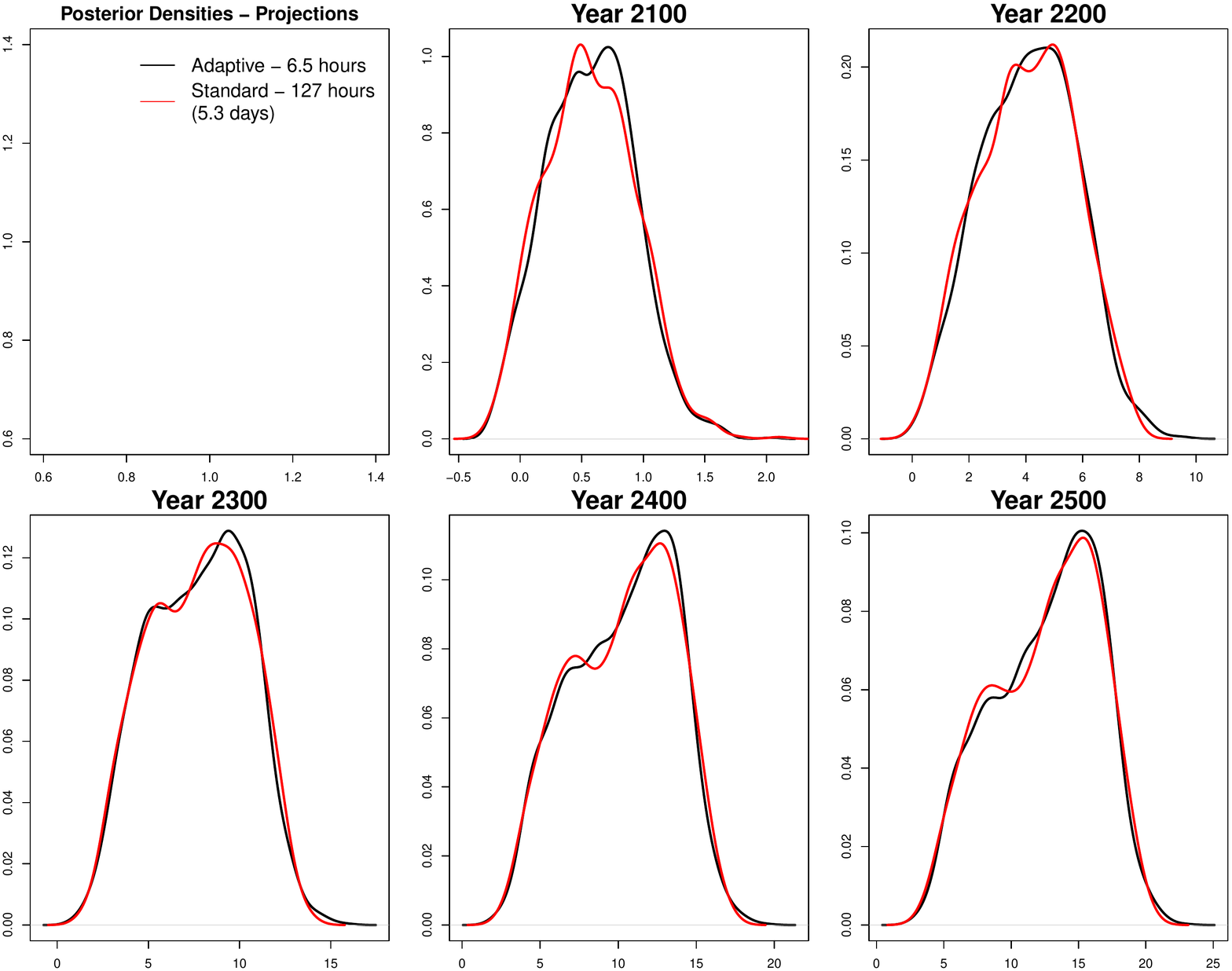}
    \caption{Posterior densities of sea level projections using adaptive particle-based approach (solid black lines) and the standard approach (dashed red lines). The adaptive particle-based approach goes through 4 cycles and runs 14 updates in the mutation stage with a total calibration wall time of 6.5 hours. The standard particle-based approach goes through 10 cycles and runs 45 updates in the mutation stage with a total calibration wall time of 127 hours (5.3 days). Posterior densities for both methods are comparable.}
\label{SuppFig:StandardAdaptiveProj}
\end{figure}

Upon examining the standard implementation, we observe that the distribution of the particles do not change after 10 Metropolis-Hastings updates of the mutation stage. Therefore, the remaining 35 Metropolis-Hastings updates are redundant. Moreover, the posterior densities of the model parameters (Figure \ref{SuppFig:StandardAdaptivePar}), observational records (Figure \ref{SuppFig:StandardAdaptiveObs}), and sea level projections (Figure \ref{SuppFig:StandardAdaptiveProj}) for both methods (standard vs. adaptive) are very much similar.

\section{Fundamental Equations for the PSU3D-ICE Model}
In the main paper, the ice-sheet model is treated as a 'black box' within our calibration framework. To provide an overall picture of the physical ice-sheet model, we presents its main equations. In a sense, they are its most fundamental partial and ordinary differential equations used to time-step the state of the ice sheet forward in time. Other equations, mostly parameterizations of local processes, are also used but are not as fundamental in the sense mentioned above. 

The basic aspects of continental ice sheets and models are as follows. Ice cover on continental scales forms a dome, several kilometers thick in central regions and sloping downward to its margins at much lower elevations. Thickening due to annual snowfall (which compacts to ice) in interior regions is balanced by ice velocities towards the margins, as the ice deforms slowly under its own weight. Ice is lost mostly near the margins by surface melt, basal melt, oceanic melt, and calving of marginal vertical ice faces. If the ice reaches the ocean, it can flow across the grounding line (where the bed is below sea level and ice becomes afloat), and form floating ice shelves with thicknesses of 100's m and extents of 100's km. Horizontal ice velocities are $\sim$1 to 10 meters/years in much of the central interior, increasing to 
$\sim$100's to $\sim$1000 meters/year in marginal ice streams and shelves \citep{rignot2011ice}. 

Numerical ice-sheet models predict the time-evolving ice thicknesses and temperature distributions, changing due to velocity advection and the local accumulation and ablation processes mentioned above. Ice flow is treated as a non-linear viscous fluid using scaled (simplified) equations, separately for horizontal stretching and for the vertical shear of horizontal velocities. Slow depression and rebound of the bedrock beneath the changing ice load is also modeled, as this affects ice surface elevations and ocean depths at grounding lines. These basic aspects are common to many large-scale ice-sheet models, and are described in detail in \citet{pollard2012description} and \citet{pollard2015potential}.\\

\textbf{I. Ice Thickness}
\begin{equation*}\label{Eq:IceThickness}
     \frac{\partial h}{\partial t}+\frac{\partial (\bar{u}h)}{\partial x}+\frac{\partial (\bar{v}h)}{\partial y} =\mbox{SMB}-\mbox{BMB}-\mbox{OMB}-\mbox{CMB}-\mbox{FMB},
     \end{equation*}
\noindent where $h$ is ice thickness, $\bar{u}$ is the mean horizontal ice velocity in the $x$ direction, $\bar{v}$ is the mean horizontal ice velocity in the $y$ direction, SMB is the surface mass balance, BMB is the basal melting (if grounded), OMB is the oceanic sub-ice melting or freezing (if floating), CMB is the calving loss (floating edge), and FMB is the face melt loss (floating or tidewater vertical face).
\newline

\textbf{II. Velocity Stretching:}\\
\begin{equation}\label{Eq:VelocityStretchingX}
     \frac{\partial}{\partial x}\Big[ \frac{h}{A\sigma^{n-1}} \Big( 2\frac{\partial\bar{u}}{\partial x} + \frac{\partial \bar{v}}{\partial y} \Big) \Big] + \frac{\partial}{\partial y}\Big[ \frac{h}{2A\sigma^{n-1}} \Big( \frac{\partial\bar{u}}{\partial y} + \frac{\partial \bar{v}}{\partial x} \Big) \Big] = \rho_{i}gh\frac{\partial h_{s}}{\partial x}+\frac{1}{C'^{1/m}}\frac{1}{|u_{b}|^{1-\frac{1}{m}}}u_{b},
\end{equation}
\begin{equation}\label{Eq:VelocityStretchingY}
     \frac{\partial}{\partial y}\Big[ \frac{h}{A\sigma^{n-1}} \Big( 2\frac{\partial\bar{v}}{\partial y} + \frac{\partial \bar{u}}{\partial x} \Big) \Big] + \frac{\partial}{\partial x}\Big[ \frac{h}{2A\sigma^{n-1}} \Big( \frac{\partial\bar{u}}{\partial y} + \frac{\partial \bar{v}}{\partial x} \Big) \Big] = \rho_{i}gh\frac{\partial h_{s}}{\partial y}+\frac{1}{C'^{1/m}}\frac{1}{|v_{b}|^{1-\frac{1}{m}}}v_{b},
\end{equation}
where $\bar{u}=\bar{u}_{i}+u_{b}$ and  $\bar{v}=\bar{v}_{i}+v_{b}$. Here, $\bar{u_i}$ is mean horizontal velocity from vertical shearing, and $u_b$ is basal sliding velocity in the $x$ direction. Similarly, $\bar{v_i}$ is mean horizontal velocity from vertical shearing, and $v_b$ is basal sliding velocity in the $y$ direction. $u_i$ is the horizontal velocity in the x direction from vertical shearing (i.e., minus its value at the base), and $v_i$ is the horizontal velocity in the y direction from vertical shearing. $A$ is the ice rheological coefficient, $\sigma$ is the effective stress (second invariant of the stress tensor), $n=3$ is the ice rheological exponent, and $g$ is gravitational acceleration. $C'$ is the basal sliding coefficient between bed and ice and $m$ is the basal sliding exponent. $h_s$ is ice surface elevation, where 
$$
h_{s} = \left\{ \begin{array}{cl}
 h+h_{b} &\mbox{, if grounded} \\
  (\frac{p_{w}-p_{i}}{p_{i}})h &\mbox{, if floating,}
       \end{array} \right.
$$

\noindent where $h_b$ is the bedrock elevation, $\rho_{w}$ is the ocean water density, and $\rho_{i}$
is ice density. 
\newline

\textbf{III. Velocity Shearing:}\\
\begin{equation*}\label{Eq:VelocityShearingX}
     \frac{\partial u_{i}}{\partial z}=-2A\sigma^{n-1}\Big( \rho_{i}gh\frac{\partial h_{s}}{\partial x}-L_{x} \Big)\times \Big( \frac{h_{s}-z}{h} \Big),
\end{equation*}
\begin{equation*}\label{Eq:VelocityShearingY}
     \frac{\partial v_{i}}{\partial z}=-2A\sigma^{n-1}\Big( \rho_{i}gh\frac{\partial h_{s}}{\partial y}-L_{y} \Big)\times \Big( \frac{h_{s}-z}{h} \Big),
\end{equation*}
where $z$ is the vertical elevation, $L_x$ is the left hand side of Equation \ref{Eq:VelocityStretchingX}, and $L_y$ is the left hand side of Equation \ref{Eq:VelocityStretchingY}.
\newline

\textbf{IV. Temperature:}\\
The prognostic equation for internal ice temperatures
$T$ is
\begin{equation*}\label{Eq:Temperature}
     \frac{\partial T}{\partial t}+u\frac{\partial T}{\partial x}+v\frac{\partial T}{\partial y}+w\frac{\partial T}{\partial z} =\frac{1}{\rho_{i}c_{i}}\frac{\partial }{\partial z}\Big(\frac{k\partial T}{\partial z}\Big) + \frac{Q}{\rho_{i}c_{i}},
     \end{equation*}
where $u=u_{b}+u_{i}(z)$, $v=v_{b}+v_{i}(z)$, and $w$ is deduced from continuity. $k$ is the ice thermal conductivity, $Q$ is internal deformational heating, and $c_{i}$ is the specific heat of ice.
\newline

\textbf{V. Bedrock Elevation:}\\
The rate of change of bedrock elevation is given by:
\begin{equation*}\label{Eq:BedrockRate}
    \frac{\partial h_{b}}{\partial t} = -\frac{1}{\tau}(h_{b}-h_{b}^{eq}+w_{b}), 
     \end{equation*}

\noindent where $h_{b}^{eq}$ is its equilibrium value and $\tau= 3000$ years is the asthenospheric isostatic relation time scale. The downward deflection of the fully relaxed response (as if the asthenosphere had no lag), $w_{b}$, is given by:
\begin{equation*}\label{Eq:Deflection}
     D\nabla^{4}w_{b} + \rho_{b}gw_{b} = q,
     \end{equation*}
where $D$ is the flexural rigidity of the lithosphere, $\rho_{b}$ is the bedrock (asthenospheric) density, and the applied load $q$ is:
\begin{equation*}\label{Eq:AppliedLoad}
     q=\rho_{i}g(h-h^{eq})+ \rho_{w}g(h_{w}-h_{w}^{eq}),
     \end{equation*}
where $h_{w}$ is ocean column thickness, $h_{w}^{eq}$ is ocean column thickness in the equilibrium state, and $h^{eq}$ is ice thickness in the equilibrium state.
\bibliography{references.bib}

